\documentclass[epj]{svjour}
\usepackage{latexsym,graphicx,epsfig,psfrag,here,color,citesort}

\begin{document}

\newcommand{\eq}{\begin{eqnarray}}
\newcommand{\en}{\end{eqnarray}}    
\renewcommand{\theequation}{\arabic{section}.\arabic{equation}}
\newcommand{\mathbold}[1]{\mbox{\boldmath $\bf#1$}}
\newcommand{\GeV}{\mbox{GeV}}
\newcommand{\MeV}{\mbox{MeV}}

\title{The pion-nucleon scattering lengths from pionic deuterium
\thanks{This research is part of the EU Integrated Infrastructure 
Initiative Hadron Physics Project under contract number RII3-CT-2004-506078. 
Work supported in part by DFG 
(SFB/TR 16, ``Subnuclear Structure of Matter'').}}
\author{Ulf-G. Mei{\ss}ner\inst{1,2} \and Udit Raha\inst{1} \and 
 Akaki Rusetsky\inst{1,3}}
\institute{
Universit\"{a}t Bonn, Helmholtz-Institut f\"{u}r Strahlen- und 
Kernphysik (Theorie),\\ 
Nu\ss allee 14-16, D-53115 Bonn, Germany
\and 
Forschungszentrum J\"{u}lich, Institut f\" {u}r Kenphysik (Theorie),
D-52425 J\"{u}lich, Germany
\and
On leave of absence from: High Energy Physics Institute,
Tbilisi State University,\\
University St.~9, 380086 Tbilisi, Georgia
}
\date{Received: date / Revised version: date}
% The correct dates will be entered by Springer
%
\abstract{
We use the framework of effective field theories  to 
discuss the determination
of the $S$-wave $\pi N$ scattering lengths from the recent high-precision measurements
of pionic deuterium observables. The theoretical analysis proceeds in 
several steps. Initially, the precise value of the pion-deuteron 
scattering length $a_{\pi d}$ is extracted from the data. Next,
$a_{\pi d}$ is related to the $S$-wave $\pi N$ scattering lengths $a_+$ and 
$a_-$. We discuss the use of
 this information for constraining the values of these 
scattering lengths in the full analysis, which also includes the input from the
pionic hydrogen energy shift and width measurements, and
throughly investigate the accuracy limits for this procedure.
In this paper, we also give
a detailed comparison to other effective field theory approaches,
as well as with the earlier work on the
subject, carried out within the potential model and multiple scattering 
framework.
 % end of PACS codes
} %end of abstract
\maketitle

\newpage

\setcounter{equation}{0}
\section{Introduction}
\label{sec:intro}

The Pionic Hydrogen collaboration at PSI \cite{PSI1,PSI2} 
has performed high precision measurements of the strong interaction shift
$\varepsilon_{1s}$ and width $\Gamma_{1s}$ of the $1s$ state of pionic
deuterium from the $3p-1s$ $X$-ray transition. The (complex) pion-deuteron
scattering length was extracted from these measurements with the use of
the leading-order Deser formula \cite{Deser} 
\eq\label{eq:Deser}
-\varepsilon_{1s}+i\,\frac{\Gamma_{1s}}{2}=\frac{4E_{1s}}{r_B}\, a_{\pi d}\, ,
\en
where $E_{1s}=\frac{1}{2}\,\alpha^2\mu_d$ is the Coulomb binding 
energy in the $1s$ state, and $r_B=(\alpha\mu_d)^{-1}$ denotes the Bohr radius
(in these formulae, $\mu_d$ stands for the reduced mass of the $\pi d$ 
system).
The most recent measurement of the pion-deuteron scattering length by the 
Pionic Hydrogen collaboration at PSI \cite{PSI2} yields
\eq\label{eq:a-pid}
a_{\pi d}=\biggl( -0.0261\,(\pm 0.0005)
+i\, 0.0063\,(\pm 0.0007)\biggr) M_\pi^{-1}\, .
\nonumber\\
\en
Performing experiments to determine
$a_{\pi d}$ is usually justified by the possibility of extracting independent
information abo\-ut the $\pi N$ $S$-wave isoscalar ($a_+$) and isovector ($a_-$)
scattering lengths. 
What makes this enterprise particularly interesting is the fact that in the
multiple scattering theory $a_{\pi d}={\rm const}\cdot a_++{\rm correction~terms}$. 
If one could accurately evaluate the higher-order terms in this
expression, then a precise measurement of $a_{\pi d}$ would enable
one to constrain the value of $a_+$, which is in general a rather delicate task.
The reason for this is that, since $a_+$
is much smaller than the isospin-odd 
scattering length $a_-$, a very high accuracy is needed in order 
to determine $a_+$ from the measurements of the 
linear combinations of $a_+$ and $a_-$. This is exactly the case
in the experiments measuring the pionic hydrogen energy shift and width,
which enable one to 
determine the combinations $(a_++a_-)$ and $a_-$, respectively. 
Thus, the measurement of $a_{\pi d}$ contributes a complementary piece
of information about the scattering lengths, which can be used in the complex
theoretical analysis finally 
aimed at the determination of both $a_+$ and $a_-$ 
\cite{PSI3}. We wish to mention here that these scattering lengths are
quantities of fundamental importance in low-energy hadronic physics,
since they test the QCD symmetries and the exact pattern of the chiral symmetry 
breaking. Moreover, since the knowledge of these scattering lengths 
places a constraint on the  $\pi N$ interactions at low energy, it also affects
our understanding of more complicated systems where $\pi N$ interaction serves
as an input, e.g. $NN$ interaction, $\pi$-nucleus scattering, three-nucleon
forces, etc.

Expressing the $\pi d$ scattering 
length in terms of the parameters characterizing the underlying pion-nucleon
and nucleon-nucleon dynamics is one of the classical problems in
conventional nuclear physics based on the potential scattering formalism,
see, e.g. \cite{Afnan,Faldt,Mizutani,Thomas,Deloff,Baru,Loiseau,Hanhart,Wilkin}. Note, however, that the experiments
on pionic deuterium will be used to extract the $\pi N$ scattering
lengths in QCD and not in any potential model. In other words, using 
the latter in order to establish the relation between $\pi d$ and $\pi N$
scattering lengths
introduces a theoretical error in the analysis of the experimental data, 
whose magnitude is very hard to control.

In recent years, the problem of a very low energy pion-deuteron scattering
has been studied within the framework of effective field theories (EFTs).
The method originates from the seminal paper of Weinberg \cite{Weinberg},
where chiral Lagrangians have been systematically applied
for the description of interactions of pions with nuclei. 
In this paper, by using the chiral Lagrangian, 
one calculates the set of diagrams contributing to the
``irreducible transition kernel'' for the pion scattering on two nucleons, and 
the result is then sandwiched between ``realistic''
deuteron wave functions in order to
evaluate the scattering amplitude of the pion on the deuteron. In actual
calculations carried out in Ref.~\cite{Weinberg}, the phenomenological deuteron
wave function for the Bonn potential model has been used. We wish to note
that this last step, in general, can be justified within the framework of the 
effective field theories, only if the particular process which one is going
to describe is dominated by the long-range mechanisms, e.g. by  one-pion exchange.
On the other hand, 
when the calculations within such a ``hybrid'' approach
are pursued in higher orders in Chiral Perturbation theory (ChPT), 
the kernels grow faster with a large momenta and probe shorter distances.
Moreover, this short-distance behavior is not necessarily correlated
to that of the phenomenological wave function.
From this we conclude that
in order to obtain a systematic description of the pion-deuteron system, 
it is preferable to use the deuteron wave 
functions and the transition kernels, evaluated within the same 
field-theoretical setting
-- in this case, no specific conjecture
about the dominance in the unobservable quantities,
like the kernel or the wave function is needed.
For the latest work within the hybrid approach, see e.g.
\cite{Kaiser,Doring}.

Further development of the approach based on chiral Lagrangians
(see, e.g. \cite{Kaiser,Doring,BBLM,Borasoy,Beane,Bernard}), has
followed different paths. In the paper \cite{Borasoy} one has used the 
framework with perturbative pions, whereas the authors of Ref.~\cite{Beane} 
make use of the so-called Heavy Pion EFT (HP EFT) with the dibaryon field.
The latter approach is quite close to the one used in the present work.
The technique used in Refs. \cite{Borasoy,Beane} 
has the advantage that one may easily
construct the deuteron wave function in a closed form, since the lowest-order
nucleon-nucleon interactions are described by a contact four-nucleon vertex.
The central problem in both the papers is related to the calculation 
of one particular diagram, describing double scattering of
pions. These calculations lead to  a very strong
scale dependence near $\mu\simeq M_\pi$ -- a natural choice of the
scale parameter in this sort of the effective theories. 
Since, on the other hand,
this dependence must be canceled by a contribution from the low-energy
constant (LEC), which describes pointlike interactions of four nucleons and
two pions, we easily conclude that the magnitude of this  LEC can not
be small. 
In the absence of any information about the actual value of this LEC apart
from naive dimensional order-of-magnitude estimates based on the naturalness arguments,
one may finally conclude that the theoretical uncertainty in the relation
of $\pi d$ and $\pi N$ scattering lengths should be very large.

On the other hand, the results obtained in Ref.~\cite{Bernard} 
seem not to be in agreement with those of
Refs. \cite{Borasoy,Beane}. The method which is used
in Ref.~\cite{Bernard}, is a systematic extension and elaboration
of  Weinberg's original
proposal, where both the transition kernel and the deuteron wave function are
constructed in ChPT (note also, that the systematic derivation of the unitary and the 
energy-in\-de\-pen\-dent potentials within this framework has been discussed 
recently in Ref.~\cite{Krebs}). The approach uses cutoff regularization to 
deal with
the potentials that are growing for a large three-momenta. The typical
scales for the cutoff mass $\Lambda$ are somewhat smaller than the
hadronic scale in QCD $\sim 1~{\rm GeV}$ (depending on the order in ChPT 
in which the calculations are carried out). The results of the calculations
are $\Lambda$-dependent, which is a reminiscent of the scale-dependence
in the dimensional regularization scheme. The bulk of this $\Lambda$-dependence
should be canceled by an analogous dependence in the LECs, and the remainder,
which is an artifact of the non-perturbative formalism used, should be of a 
higher order in ChPT. In Ref.~\cite{Bernard}, the cutoff dependence of the 
$\pi d$ scattering length has been studied, with the LECs assumed to 
vanish. In a remarkable contrast with Refs.
\cite{Borasoy,Beane}, the $\Lambda$-dependence of the results in 
Ref.~\cite{Bernard} turns out to be very mild, 
thereby concluding that the LECs must have a weak cutoff
dependence. If one could interpret the cutoff dependence as an estimate
of the uncertainty of the method, then the results of Ref.~\cite{Bernard} 
would amount to a rather accurate prediction of the 
$\pi d$ scattering length within the framework of ChPT.

The present situation which was described above is unacceptable from
the point of view of both the theory and the phenomenological analysis of the
data. From the theoretical point of view, the calculations  carried out in
Ref.~\cite{Bernard}, clearly indicate that the diagrams in which the virtual
pions are emitted or annihilated, are strongly suppressed. 
This phenomenon originates from the infrared enhancement of a certain 
class of the diagrams in the Weinberg scheme, as well as the 
threshold suppression of the diagrams containing pseudoscalar vertices.
In order to accommodate the above feature in the theory with 
non-perturbative pions, in Ref.~\cite{Bernard}
a novel counting, inspired by the HP EFT,
has been enforced on top of the conventional ChPT
Lagrangian. Stated differently, this means that simpler effective
theories, which were used in Refs. \cite{Borasoy,Beane}, are  
physically adequate for the problem considered. How can it then be that
using a simpler theory, we get an answer which contradicts
the answer obtained in Weinberg's framework \cite{Bernard}, the very 
approach one starts from? From the point of view of 
phenomenology, the existing conflicting predictions, on the one hand,
do not encourage the 
experimentalist's efforts to measure the pion-deuteron scattering length to
a better accuracy and on the other hand, suggest that the values
of the $\pi N$ scattering lengths extracted from the analysis of the $\pi d$
data should be taken with a grain of salt.

The aim of the present paper is to perform a thorough investigation
of pion-deuteron scattering at threshold within the framework of 
low-energy effective theories. In particular, we plan to clearly establish
the limits of accuracy for extracting $\pi N$ scattering lengths from the
measured $\pi d$ scattering lengths. We also perform a detailed
study of the above-mentioned discrepancy between the results obtained
within the HP EFT and in the Weinberg approach.
Moreover, the investigation of this subtle question, in our opinion,
is by itself very informative and sheds light on many peculiar aspects of 
the effective field theories in general. 

\begin{sloppypar}
The complex problem, which we are going to consider in this paper, naturally 
falls into several sub-problems, which are characterized by distinct momentum
scales. Consequently, instead of trying to describe everything at once,
it is convenient to construct a tower of effective field theories, matched
one to another, each designed for one particular momentum scale.
\end{sloppypar}
\begin{itemize}
\item[i)]
At the momentum scales $\alpha\mu_d \simeq 1~{\rm MeV}$, the charged 
pion and the deuteron form an atom, whose observables are measured by the 
experiment. The characteristic distances in such an atom -- hundreds of fm --
are much larger than the deuteron size, and the binding energy in the ground
state, which almost coincides with the Coulomb binding energy
$E_{1s}=\frac{1}{2}\,\alpha^2\mu_d\simeq 3.5\cdot 10^3~{\rm eV}$, is much smaller,
than the binding energy of the deuteron $\epsilon=2.22457~{\rm MeV}$. Stated 
differently, at these energies the deuteron can not be resolved
as a composite particle, and the effective theory, which describes the atom,
contains the deuteron field (not the nucleon fields) 
as an elementary degree of freedom. The hard momentum scale in this 
effective theory is given by the average value of the three-momentum 
of the nucleons bound within the deuteron, 
$\gamma=\sqrt{\epsilon m}\simeq 45~{\rm MeV}$, where 
$m$ stands for the nucleon mass. 
The expansion parameter in this theory is given by the ratio of the scales 
$\alpha\mu_d/\gamma=O(\alpha)$.
The output from the calculations
within this effective theory is relation which connects 
the measured energy shift of the $\pi d$ bound state to the $\pi d$ scattering
amplitude at threshold in the next-to-leading order in isospin breaking. 
In its turn, the latter at the leading order in isospin breaking coincides
with the $\pi d$ scattering length $a_{\pi d}$.

\item[ii)]
Extracting the scattering length from the pionic de\-u\-te\-ron one next has to find
the relation of this quantity to the $\pi N$ $S$-wave
scattering lengths $a_+$
and $a_-$. In order to achieve this goal, we have to construct another 
effective field theory, in which the independent degrees of freedom are
the pion and the nucleon fields, whereas the deuteron emerges as a bound state of 
the proton and the neutron. The characteristic momentum scale in this theory
is defined by the binding momentum $\gamma$. Furthermore, a careful analysis 
of the results of Ref.~\cite{Bernard} provides us with an important clue: 
The processes, in which the virtual creation and annihilation of pions takes
place, are suppressed as compared to the processes where this does 
not occur (although both processes may formally have the same chiral order).
Note that these processes naturally come together in the 
conventional relativistic QFT.
This fact clearly indicates that the most economic way to describe $\pi d$
scattering at threshold is to design an effective field theory, in which 
the pion creation and annihilation processes are explicitly excluded --
all vertices in the Lagrangian contain equal number of ingoing and outgoing
pions and nucleons. 
The whole information about these processes is, however, not lost: it
is included in the pertinent LECs of such an effective theory.
Moreover, it is also clear that for such small energies, one can treat
kinematical relativistic factors as perturbations both for pions and for 
 nucleons.

\begin{sloppypar}
The calculations of the deuteron properties in the abo\-ve-des\-cri\-bed theory,
which will be referred to as the heavy-pion effective theory
hereafter,
dramatically simplify and can be performed analytically.
The output of the calculations
is the quantity $a_{\pi d}$, expressed in terms of the threshold parameters of
the $\pi N$ and $NN$ scattering. The hard scale in such a theory
is given by the pion mass $M_\pi$, and the expansion parameter
is given by the ratio of scales $\gamma/M_\pi\simeq 1/3$. The matching to the previous
effective theory is performed for the $\pi d$ scattering amplitude at 
threshold: this quantity must be the same in both theories.
\end{sloppypar}

Note also that from now on we neglect all isospin-breaking effects
(one could not do this at the earlier step, because the pionic deuterium is 
created predominately by Coulomb interactions.). In this approximation,
the threshold scattering amplitude coincides with the $\pi d$ scattering length
$a_{\pi d}$. If needed, the isospin-breaking effects can be turned on later.

\item[iii)]
The simplicity of the calculations 
in the HP EFT co\-mes at the cost of the large size of
the LECs. Since the hard scale of the theory is determined by $M_\pi$, 
this is also the scale that enters in the estimate of the size of the (unknown)
LECs in the assumption that these LECs have the natural size
(note that some LECs might be parametrically enhanced as compared to the value
which is expected on the purely dimensional grounds, see below).
On the other hand, if the calculations are done in ChPT,
the natural-size LECs are suppressed by a higher scale 
$4\pi F_\pi\simeq 1~{\rm GeV}$ rather than $M_\pi$.
Thus, the rationale
for performing calculations in the Weinberg framework can be formulated
as follows. In these calculations, one ``resolves'' the dynamics of the system
at the scales from $M_\pi$ up to the scale $\sim 1~{\rm GeV}$, which is the 
energy range
where the interactions in the system of few pions and nucleons are 
predominately determined by (multi)-pion exchanges.
One may then assume that the
bulk contribution to the HP LECs comes from the momentum region
between $M_\pi$ and $1~{\rm GeV}$ and can be expressed in terms of pion
loops, which are calculated in the Weinberg scheme.

\begin{sloppypar} 
If we suppose that 
such a scheme is realized, we arrive at the effective theory, where the 
characteristic momenta are of order $M_\pi$ and the hard scale is given
by $4\pi F_\pi$. The expansion parameter, in the absence of other scale,
is given by the ratio of two scales $M_\pi/(4\pi F_\pi)$.
The matching to the HP EFT 
is performed for the $S$-matrix element of the process $\pi NN\to \pi NN$, that
determines a particular LEC of the HP Lagrangian.
One of the objectives of the present paper is to find out
whether doing the calculations
in the Weinberg framework and performing the matching to the HP EFT 
enables one to indeed reduce the uncertainty related to the choice of LECs.
\end{sloppypar}

\end{itemize}

The organization of the paper follows the above-des\-cri\-bed scheme
of ``nested'' effective field theories. Namely,
in section \ref{sec:deuterium} we consider
the precise extraction of the $\pi d$ scattering length from the experimental
data on the pionic deuterium. Then, in section \ref{sec:NR}, we construct
the systematic heavy-pion effective theory (HP EFT)
in order to calculate the $\pi d$ 
scattering lengths in terms of the threshold parameters of $\pi N$ and
$NN$ interactions. In order to establish the connection to ChPT in the Weinberg
scheme, in section \ref{sec:ChPT} we perform the matching of the threshold
amplitudes in both theories. We also provide 
a numerical analysis and discuss the
question of accuracy. A detailed comparison to the existing approaches is
carried out in section \ref{sec:lit}. Finally, section \ref{sec:concl}
contains our conclusions.

\setcounter{equation}{0}
\section{Pionic deuterium}
\label{sec:deuterium}

In the experiment at PSI \cite{PSI1,PSI2}, one measures the energy
of $3p-1s$ $X$-ray transition, deducing the strong shift of the pionic
deuterium in the $1s$ state and the $\pi d$ scattering length from this 
measurement. At the first step, in order to obtain the strong shift, 
one has to subtract the so-called ``electromagnetic shift'' from the 
full measured value, where the former is calculated in the accuracy
that matches the experimental precision. At the next step, the $\pi d$
scattering length should be extracted from the strong shift by means of the
Deser-type formula (\ref{eq:Deser}). If required for accuracy considerations,
the latter relation can also be generalized to include next-to-leading order
isospin-breaking corrections.

To the best of our knowledge, complete calculations of the electromagnetic
shift in the pionic deuterium are not available in the literature,
except the results contained in
table 1 of Ref.~\cite{PSI1}, where different contributions are given without
a derivation. The
investigations in Ref.~\cite{Irgaziev} are not complete -- as the
authors themselves note, they do not include all isospin-breaking corrections
at next-to-leading order. In order to have a complete and transparent
field-theoretical treatment of the pionic deuterium problem at all levels,
we find it appropriate here to re-derive
the expression for the full energy shift at order $\alpha^4$, 
$\alpha^3(m_d-m_u)$, and to check (at least, numerically) 
the results given in table 1 of Ref.~\cite{PSI1}.

The method, which will be used in our calculations, is analogous to the one
applied recently to describe $\pi^+\pi^-$ \cite{Bern1,Bern2,Bern4,Schweizer}, 
$\pi^-p$ \cite{Bern3,Zemp}, $\pi K$ \cite{Schweizer} and $K^-p$ 
\cite{Bonn1} atoms. In this section, we display only the final results of the
calculations -- the necessary details are provided in appendix \ref{sec:shift}.
The full binding energy in a given stationary state of the 
pionic deuterium depends on the principal quantum
number $n$, on the orbital quantum number $l$ and on the total angular momentum
$j$. For a given $l$ (except $l=0$) the total angular momentum $j$ takes
the values $j=l-1,l,l+1$. This splitting, which is explicitly evaluated in
appendix \ref{sec:shift}, is tiny. The following averaged value is relevant
for the analysis of the experimental data
\eq\label{eq:E-average}
\overline{E}_{nl}=\frac{1}{6(l+1/2)}\,\sum_{j=l-1}^{l+1}(2j+1) E_{nlj}\, ,
\en
Up to the next-to-leading order in isospin breaking, the full energy shift
of the $nl$ state can be separated in what is called ``electromagnetic'' and
``strong`` parts. In order to simplify the comparison to the existing results,
the former is additionally split by hand in different pieces. Finally, 
at this order one obtains
\eq\label{eq:separation}
\overline{E}_{nl}&=&E_{nl}^{\rm em}+\Delta E_{nl}^{\rm str}\, ,
\nonumber\\[2mm]
E_{nl}^{\rm em}&=&E_{nl}^{\rm KG}+\Delta E_{nl}^{\rm rel}
+\Delta E_{nl}^{\rm fin}+\Delta E_{nl}^{\rm vac}
\, ,
\en
In the above formula, we have chosen the same naming scheme as in 
Ref.~\cite{PSI1}. Note that in this paper individual contributions
are not specified explicitly, so the identification, which is given below
and in table \ref{tab:table_em}, is performed by analogy with the pionic
hydrogen case \cite{Sigg}. Our explicit expressions are given below
\eq\label{eq:em}
E_{nl}^{\rm KG}&=&-\frac{\alpha^2\mu_d}{2n^2}\,\biggl(
1+\frac{\alpha^2}{n^2}\biggl[\frac{2n}{2l+1}-\frac{3}{4}
\biggr]\biggr)\, ,
\nonumber\\[2mm]
\Delta E_{nl}^{\rm rel}&=&\frac{3\alpha^4\mu_d^2}{8n^4(M_d+M_\pi)}\,
\biggl(\frac{4n}{l+1/2}-3\biggr)
\nonumber\\[2mm]
&-&\frac{\alpha^4\mu_d^3}{4M_dM_\pi n^4}\,
\biggl(-4n\delta_{l0}-4+\frac{6n}{l+1/2}\biggr)
\nonumber\\[2mm]
\Delta E_{nl}^{\rm fin}&=&\delta_{l0}\,
\frac{2}{3n^3}\,\alpha^4\mu_d^3(\langle r_d^2\rangle+\langle r_\pi^2\rangle)\, ,
\en
where $M_d$ denotes the mass of the deuteron. An explicit expression for the vacuum polarization contribution is gi\-ven in Ref.~\cite{Eiras}, see Eq.~(3) 
of that paper.

In order to be able to compare with the existing results, 
our numerical calculations have been performed for the same values
of the input parameters as in Ref.~\cite{PSI1}.
We take the deuteron binding energy to be $\epsilon=2.22457~{\rm MeV}$,
and the charge radii of the deuteron and of the pion are
taken to be equal $\langle r_\pi^2\rangle^{1/2}=0.657~{\rm fm}$
and $\langle r_d^2\rangle^{1/2}=2.106~{\rm fm}$, respectively. 
The calculations were performed for the value of the charged pion 
mass $M_\pi=139.57018~{\rm MeV}$. 
In addition, the calculation of the finite-size correction has been performed
by using the latest data for the
charge radii $\langle r_\pi^2\rangle=0.452\pm 0.013~{\rm fm}^2$ \cite{new1}
and $\langle r_d^2\rangle^{1/2}=2.1303(66)~{\rm fm}$, see \cite{new2}
and references therein (the result changes slightly).
The results of our calculations and the 
comparison to the results of Ref.~\cite{PSI1} are given in table 
\ref{tab:table_em}. Note that we have not calculated higher-order 
(next-to-next-to-leading) 
isospin-breaking cor\-rec\-ti\-ons that are given in the last two entries of this
table. The results of the calculations from Ref.~\cite{PSI1} in these cases
should be taken at face value. It can be immediately seen  from the table that
our calculations completely confirm the results of Ref.~\cite{PSI1}
at next-to-leading order -- the agreement between the two columns is perfect.

After having subtracted the calculated electromagnetic contributions from 
the measured transition energy, one finally arrives at the strong shift, 
which is related to the $\pi d$ scattering length. Since in the $p$-states 
the strong shift is proportional to $\alpha^5$ and is thus tiny, the 
measurement of the quantity $E_{3p}-E_{1s}$ yields directly the strong shift 
in the $1s$-state. In next-to-leading order in isospin breaking, the
strong shift for the states with $l=0$ is given by
\eq\label{eq:strong}
\Delta E_{n0}^{\rm str}&=&\varepsilon_{ns}-i\,\frac{\Gamma_{ns}}{2}=
-\frac{\alpha^3\mu_d^3}{2\pi M_\pi n^3}\,
{\cal T}_{\pi d}
\nonumber\\[2mm]
&\times&\biggl\{1-\frac{\alpha\mu_d^2}{4\pi M_\pi}\,
{\cal T}_{\pi d}(s_n(\alpha)+2\pi i)+\delta_n^{\rm vac}\biggr\}\, ,
\nonumber\\[2mm]
s_n(\alpha)&=&2\biggl(\psi(n)-\psi(1)-\frac{1}{n}+\ln\alpha-\ln n\biggr)\, ,
\nonumber\\[2mm]
\psi(x)&=&\Gamma'(x)/\Gamma(x)\, ,
\en
where the quantity ${\cal T}_{\pi d}$ denotes the threshold scattering 
amplitude in the presence of photons, which is obtained from 
the conventional amplitude by subtracting all singular contributions
at threshold (see \cite{Bern1,Bern2,Bern4,Schweizer,Bern3,Zemp,Bonn1}
for more details and definitions).
The normalization of this quantity is chosen so
that in the absence of the isospin-breaking effects, it reduces to the
$\pi d$ scattering length 
\eq\label{eq:scl}
{\cal T}_{\pi d}=4\pi\biggl(1+\frac{M_\pi}{M_d}\biggr)a_{\pi d}+\cdots\, ,
\en
where the ellipses stand for terms vanishing at $\alpha=0$ and
$m_d=m_u$. These terms can be in principle evaluated in ChPT in a systematic
manner, in analogy with more simple cases of $\pi N$ 
\cite{Meissner_piN,Mojzis} and $NN$ \cite{Meissner_NN} scattering.
Further, the quantity $\delta_n^{\rm vac}
=2\delta\Psi_n(0)/\Psi_n(0)$ stands for the vacuum polarization correction to
the strong shift ($\delta\Psi_n(0)$ denotes the correction of the 
Coulomb wave function $\Psi_n(0)$ at the origin due to the vacuum polarization
effects). This correction was evaluated in Ref.~\cite{Eiras} only 
for the ground state. However, the approach used in this paper  can be
straightforwardly generalized for the radially excited states. 
Finally, it is interesting to note that the $n$-dependence 
of the correction term in
Eq.~(\ref{eq:strong}) is universal, since short-range effects are the same
in all atomic states. For this reason, even potential models (see, e.g.
\cite{Ericson:2004ps}) agree with our result in what concerns
the difference of the correction terms in the states with a different $n$.

\begin{table}[t]
\def\arraytretch{2}
\begin{center}
\begin{tabular}{lcc}
\hline\noalign{\smallskip}
Calculated corrections & ~Ref.~\cite{PSI1}~ & ~This work~ \\
to $E_{3p}-E_{1s}$~[eV] & & \\
\noalign{\smallskip}\hline\noalign{\smallskip}
Point nucleus (Klein-Gordon) & $3074.69$& $3074.69$\\
Nuclear and $\pi^-$ finite size & $-0.51$ & $-0.52$\\
& & $-0.53$  \cite{new1,new2} \\
Vacuum polarization $\alpha(Z\alpha)$& $3.72$ & $3.72$ \\
Relativistic recoil & $-0.02$ & $-0.02$ \\
Higher order radiative  & $0.04$ & -- \\
corrections & & \\
Nuclear polarization & $0.03$ & -- \\
\noalign{\smallskip}\hline
\end{tabular}

\end{center}
\caption{Comparison of the electromagnetic corrections to the $E_{3p}-E_{1s}$
transition energies, calculated in Ref.~\cite{PSI1} and in the present paper.
The second entry for the finite-size effect has been obtained, using the
latest experimental data on the charge radii.
We did not address the calculation of the last two entries in this table.}
\label{tab:table_em}
\end{table}

To summarize, in this section we have checked the validity of the procedure
which is used for the theoretical analysis of the pionic deuterium data at PSI.
The calculated electromagnetic shift agrees very well to the one given in 
Ref.~\cite{PSI1}. Further,
we have obtained the general expression
for the (complex) strong energy shift of the pionic deuterium in the
next-to-leading order in isospin breaking, in terms of the $\pi d$ threshold 
scattering amplitude. This relation should in principle be used to replace
the lowest-order formula (\ref{eq:Deser}) in the data analysis.
Note however, that the Coulomb correction which
is explicitly displayed in Eq.~(\ref{eq:strong}) (second term in the brackets),
 is of order of $2\cdot 10^{-3}$, if
one replaces ${\cal T}_{\pi d}$ by Eq.~(\ref{eq:scl}) and uses the value of
the scattering length given in Eq.~(\ref{eq:a-pid}). Note also that 
$10^{-2}\cdots 10^{-3}$ is an expected
order
of magnitude for the vacuum polarization contribution in Eq.~(\ref{eq:strong}),
see Ref.~\cite{Eiras}. Since there are no obvious reasons for having an 
anomalously large isospin-breaking correction in the quantity 
${\cal T}_{\pi d}$ either (see e.g. \cite{PSI1,PSI2} and references therein), 
in the following we do not consider 
isospin-breaking corrections to the energy shift at all and concentrate on
the lowest-order relation Eq.~(\ref{eq:Deser}). If it turns out that the 
determination of the $\pi N$ scattering lengths from the analysis of the pionic
deuterium data can be performed at a few percent level
that requires the inclusion
of the isospin-breaking effects in Eq.~(\ref{eq:Deser}), one can always go back to 
the relation Eq.~(\ref{eq:strong}).

\setcounter{equation}{0}
\section{Heavy-pion effective theory for $\mathbold{\pi d}$ scattering}
\label{sec:NR}

\subsection{The Lagrangian}

\begin{figure}[t]
\begin{center}
\includegraphics[width=7.5cm]{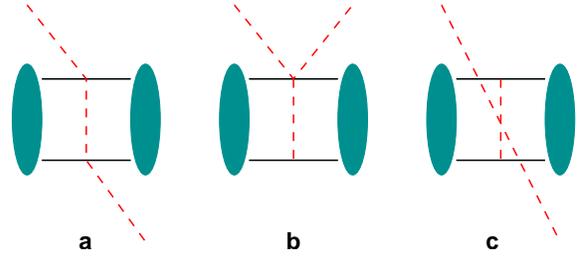}
\end{center}
\caption{Leading-order three-body graphs which contribute to the $\pi d$ 
scattering length in the Weinberg scheme (same as in Ref.~\cite{Bernard}). 
Shaded blobs stand for the deuteron
wave function, and the solid and dashed lines denote nucleons and pions, 
respectively. Numerically, graphs b+c are 
suppressed by two orders of magnitude as compared to the graph a.}
\label{fig:Bernard}
\end{figure}

The findings of Ref.~\cite{Bernard}, as well as the earlier work on the 
subject (see, e.g. \cite{Weinberg}) serve as a clear indication of the fact
that the chiral counting is not the most suitable one to be applied for the
description of low-energy $\pi d$ scattering. Most straightforwardly, this
can be visualized by comparing the contributions from the individual diagrams,
depicted in Fig.~2 of \cite{Bernard}, which is reproduced here, in 
Fig.~\ref{fig:Bernard}. The contribution from the diagrams
\ref{fig:Bernard}b+\ref{fig:Bernard}c is by two orders of magnitude smaller than the contribution from \ref{fig:Bernard}a, 
although all three diagrams emerge at the same chiral order. The reason
for this difference is that, in contrary to \ref{fig:Bernard}a, the diagrams \ref{fig:Bernard}b+\ref{fig:Bernard}c
describe processes with the virtual pion emission/absorption
(additional suppression at a small momenta is caused by the presence of the 
$\gamma_5$-vertices in the diagrams \ref{fig:Bernard}b and \ref{fig:Bernard}c).

The above discussions lead to the conclusion that it will be convenient
to describe the $\pi d$ scattering at threshold in a framework in which 
the absorption and emission of hadrons does not appear explicitly
at the level of Feynman diagrams, but is included in the couplings of
the effective Lagrangian. In this manner, we arrive at the theory
which must be in a spirit similar to
HP EFT (see \cite{Beane} and references therein).  
Below we dwell on some differences which exist between the approach
used here and in Ref.~\cite{Beane}.

\begin{itemize}
\item[i)]
The HP EFT of Ref.~\cite{Beane} uses the notion of the di\-ba\-ry\-on field, whereas
we work in terms of the elementary nucleon constituents and sum up all 
interactions in the $NN$ subsystems. After substituting the expression
for the couplings of the Lagrangian in terms of the observables (coefficients
in the effective range expansion), the ``deuteron propagator'' in the present
paper coincides with the dibaryon propagator of Ref.~\cite{Beane}
in the limit of vanishing effective range.
We opt to work in terms of nucleon field in order to make
the comparison with ChPT in the Weinberg picture more transparent.

\item[ii)]
It has been argued that the technique based on the introduction of the 
dibaryon field enables one to effectively sum up all potentially large
contributions coming from the large scattering length and the effective
range, whereas higher-order terms can be treated perturbatively.
The results of the present paper are obtained under an additional 
assumption that the ef\-fec\-ti\-ve-range term is small, leading to some technical
simplifications. This assumption is, however, not critical -- the inclusion
of the effective-range term is straightforward in our approach and does not
affect the conclusions.

\item[iii)]
In Ref.~\cite{Beane} the authors have studied one particular diagram
that corresponds to the double-scattering contribution the the $\pi d$
scattering length. In this paper, a systematic expansion of the quantity
$a_{\pi d}$ is performed in the small parameter $x=\gamma/M_\pi$, 
up to and including terms of order $x$. At this order, there are
additional diagrams apart from the one mentioned above.

\item[iv)]
The most important question is the convergence of the series for the 
$\pi d$ scattering length. We believe
that there are (indirect) indications which testify in favor of the
convergence. First of all, at the momenta $\gamma\simeq 45~{\rm MeV}$,
the pionless effective theory gives still a reasonable description of the
$NN$ sector. As was mentioned, this fact is in agreement with the observation 
made in Ref.~\cite{Bernard} that the ``modified power counting'' in the 
$\pi d$ scattering length works much better than the original chiral 
counting\footnote{
And {\it vice versa}, one may treat the HP EFT, as the systematic 
field-theoretical realization of the counting $\gamma\ll M_\pi\ll 4\pi F_\pi$, 
which is heuristically implemented in the ``modified power counting''
of Ref.~\cite{Bernard}.}. Yet another justification of the
method is provided by the well-known fact that in the Faddeev approach,
the multiple-scattering series
for the threshold $\pi d$ scattering amplitude are rapidly
convergent, since the $\pi N$ scattering lengths are much smaller than the
deuteron radius (see e.g. \cite{Doring} and references therein).

\end{itemize}

After these preliminary remarks, let us consider the Lagrangian of our theory.
By construction, the  Lagrangian does contains only vertices
with the same number of the incoming and outgoing pions and nucleons.
Restricting ourselves to the non-derivative couplings, one may easily write
down the most general form of this Lagrangian (in addition, we omit below
also all three-body non-derivative terms which contain $\pi_+$, $\pi_0$ and/or the
$NN$-pair in the $^1S_0$ state: such terms 
do not contribute to the $\pi^- d$ threshold scattering amplitude 
at the accuracy we are working)
\eq\label{eq:L_HP}
{\cal L}&\!\!=\!\!&\mathbold{\pi}^\dagger\biggl(i\partial_t-M_\pi
+\frac{\triangle}{2M_\pi}+\frac{\triangle^2}{8M_\pi^3}+\cdots\biggr)
\mathbold{\pi}
\nonumber\\[2mm]
&\!\!+\!\!&\psi^\dagger\biggl(i\partial_t-m+\frac{\triangle}{2m}
+\frac{\triangle^2}{8m^3}+\cdots\biggr)\psi
\nonumber\\[2mm]
&\!\!+\!\!&\psi^\dagger(d_+(\mathbold{\pi}^\dagger\mathbold{\pi})
+id_-(\mathbold{\pi}^\dagger\times\mathbold{\pi}))\psi
\nonumber\\[2mm]
&\!\!+\!\!&c_0(\psi^TP_a\psi)^\dagger(\psi^TP_a\psi)
+c_1(\psi^TP_i\psi)^\dagger(\psi^TP_i\psi)
\nonumber\\[2mm]
&\!\!+\!\!&f_0(\psi^TP_i\psi)^\dagger(\psi^TP_i\psi)\pi_-^\dagger\pi_-+\cdots\, ,
\en
where the ellipses stand for the omitted three-body terms, as well as for
the higher-order terms in the derivative expansion. The non-relativistic
pion and nucleon fields are defined as
$\mathbold{\pi}=(\pi_1,\pi_2,\pi_3)$, where $\sqrt{2}\pi_\pm=\pi_1\mp i\pi_2$,
$\pi_0=\pi_3$ and $\psi=\begin{pmatrix}{p\cr n}\end{pmatrix}$.
Further, $P_a$ and $P_i$ denote the
projection operators onto the $^1S_0$ and $^3S_1$ states, respectively
\eq\label{eq:Projectors}
P_a=\frac{1}{\sqrt{8}}\,\tau_2\tau_a\sigma_2\, ,\quad\quad
P_i=\frac{1}{\sqrt{8}}\,\sigma_2\sigma_i\tau_2\, ,
\en
where $\sigma$ and $\tau$ are the Pauli matrices in the spin and isospin 
space, respectively. Note that we have not introduced an elementary deuteron
field in the Lagrangian. In our approach, the deuteron emerges as a bound-state
pole in the Green functions after the non-perturbative resummation of the 
lowest-order four-nucleon vertex.

In the above Lagrangian, $d_{\pm},c_{0,1},f_0$ stand for the effective
low-energy couplings. These should be determined from matching of the 
various observables. Using  dimensional regularization for calculating
the loops enormously simplifies these calculations: 
as it is well known, all two-particle bubbles vanish at threshold and the
results of the tree-level matching in the two-particle sectors remain
intact. For example, the constants $d_{\pm}$ are related to the $\pi N$ 
scattering lengths through
\eq\label{eq:apm}
a_\pm=\frac{mM_\pi}{2\pi(m+M_\pi)}\, d_\pm\, ,
\en
and this relation remains unaffected by loop corrections.
As concerning the constant $c_1$,
we find it more convenient to perform the matching in the $^3S_1$
channel for the deuteron binding energy, and not for the scattering length
in the $np$ system. The difference between these two methods shows up at
higher orders.

\subsection{The deuteron}

In the two-nucleon sector of the HP EFT, there is
no trace of pion-nucleon interactions: $NN$ scattering is described completely
in terms of contact four-nucleon interactions. The only possible loop diagrams
are the $s$-channel bubbles containing the vertices with the couplings
$c_{0,1}$. At higher orders, one should also include the derivative 
four-nucleon vertices.

Consider the following connected four-point function in $D$ dimensions
\eq\label{eq:4P}
&&(2\pi)^D\delta^D(p_1+p_2-q_1-q_2)G_N(P;p,q)
\nonumber\\[2mm]
&\!\!=\!\!&
i^4\int d^Dx_1d^Dx_2d^Dy_1d^Dy_2\,
{\rm e}^{ip_1x_1+ip_2x_2-iq_1y_1-iq_2y_2}
\nonumber\\[2mm]
&\!\!\times\!\!&
\langle 0|T\psi(x_1)\psi(x_2)\psi^\dagger(y_1)\psi^\dagger(y_2)|0\rangle_c\, ,
\en
where all spin and isospin indices have been suppressed, and the CM and 
relative momenta are defined as
\eq\label{eq:CM}
&&P=p_1+p_2=q_1+q_2\, ,
\nonumber\\[2mm]
&&p=\frac{1}{2}(p_1-p_2)\, ,\quad\quad
q=\frac{1}{2}(q_1-q_2)\, .
\en
The center-of-mass (CM) 
frame corresponds to\\ $P^\mu=(P^0,{\bf 0})$, and $D\to 4$ in physical space.

At the energy $P^0\to P_B^0({\bf P})=M_d+{\bf P}^2/2M_d$, 
the four-point function Eq.~(\ref{eq:4P})
develops a bound-state pole corresponding to the deuteron
\eq\label{eq:Bpole}
G_N(P;p,q)\rightarrow -i\,
\sum_{i=1}^3\frac{\Psi_i(p)\Psi_i^\dagger(q)}{P_B^0({\bf P})-P^0}
+\mbox{regular terms}\, ,
\nonumber\\
\en
where the deuteron wave function is defined as
\eq\label{eq:wf}
\Psi_i(p)&=&\int d^Dx{\rm e}^{ipx}\langle 0|T\psi(\frac{x}{2})
\psi(-\frac{x}{2})|B,i\rangle\, ,
\nonumber\\[2mm]
\Psi_i^\dagger(q)&=&\int d^Dy{\rm e}^{-iqy}\langle B,i|T\psi^\dagger(\frac{y}{2})
\psi^\dagger(-\frac{y}{2})|0\rangle\, ,
\en
and the sum in Eq.~(\ref{eq:Bpole}) runs over the polarizations of the 
deuteron spin.

On the other hand, one may evaluate the 4-point function Eq.~(\ref{eq:4P}) with
the use of the Lagrangian Eq.~(\ref{eq:L_HP}) that amounts to the resummation
of the geometrical series corresponding to the $s$-channel bubbles with 
four-nucleon vertices. 
Further, since at the leading order the deuteron is a 
purely $^3S_1$ state, we may put $c_0=0$ in order to get the deuteron pole. 
As the result of this resummation, one gets
\eq\label{eq:resum}
&&G_N(P;p,q)=
\sum_{i=1}^3\frac{2P_i^\dagger}{(w({\bf p}_1)-p_1^0)(w({\bf p}_2)-p_2^0)}
\nonumber\\[2mm]
&&\hspace*{.7cm}\times\,\frac{ic_1}{1-c_1J(P^0,{\bf P})}\,
\frac{2P_i}{(w({\bf q}_1)-q_1^0)(w({\bf q}_2)-q_2^0)}
\nonumber\\[2mm]
&&\hspace*{.7cm}+\,\mbox{terms with $c_0$}\, ,
\en
where, $w({\bf p})=m+{\bf p}^2/2m$, and
\eq\label{eq:JP0}
&&J(P^0,{\bf P})=\int\frac{d^Dl}{(2\pi)^Di}\,\frac{1}
{(w({\bf l})-l^0)(w({\bf P}-{\bf l})-P^0+l^0)}
\nonumber\\[2mm]
&&\hspace*{.4cm}=\frac{m^{\frac{d}{2}}\Gamma(1-d/2)}{(4\pi)^{\frac{d}{2}}}\,
\biggl(2m-P^0+\frac{{\bf P}^2}{4m}-i0\biggr)^{\frac{d}{2}-1}\, ,
\en
with $d=D-1$.
In the CM frame ${\bf P}=0$ the denominator in Eq.~(\ref{eq:resum}) develops
a pole at $P^0=M_d=2m-\epsilon$. This gives (in $d$ dimensions)
\eq\label{eq:pole}
1-c_1J(M_d,{\bf 0})=0\, ,\quad\quad
c_1=\frac{(4\pi)^{\frac{d}{2}}\epsilon^{1-\frac{d}{2}}}{m^{\frac{d}{2}}
\Gamma(1-d/2)}\, .
\en
Finally, in the CM frame the behavior of the Green function near the 
deuteron pole is given by
\eq\label{eq:resum1}
&&G_N(P;p,q)\rightarrow
\sum_{i=1}^3\frac{2P_i^\dagger}{(w({\bf p}_1)-p_1^0)(w({\bf p}_2)-p_2^0)}\,
\frac{iZ}{M_d-P^0}
\nonumber\\[2mm]
&&\hspace*{.3cm}\times\,\frac{2P_i}{(w({\bf q}_1)-q_1^0)(w({\bf q}_2)-q_2^0)}
+\mbox{regular terms}\, ,
\en
where the deuteron wave function renormalization constant is given by
\eq\label{eq:Zd}
Z=\frac{(4\pi)^{\frac{d}{2}}\epsilon^{2-\frac{d}{2}}}
{m^{\frac{d}{2}}\Gamma(2-d/2)}\, .
\en

\subsection{Pion-deuteron scattering}

The pion-deuteron scattering amplitude can be extracted from the 6-point
connected Green function
\eq\label{eq:6P}
&&\!(2\pi)^D\delta^D(p_1+p_2+p_3-q_1-q_2-q_3)\,G(P,Q;p,q;p_3,q_3)
\nonumber\\[2mm]
&&\!=i^6\int d^Dx_1d^Dx_2d^Dx_3d^Dy_1d^Dy_2d^Dy_3
\nonumber\\[2mm]
&&\!\times\,{\rm e}^{ip_1x_1+ip_2x_2+ip_3x_3-iq_1y_1-iq_2y_2-iq_3y_3}
\nonumber\\[2mm]
&&\!\times\,
\langle 0|T\psi(x_1)\psi(x_2)\pi_-(x_3)
\psi^\dagger(y_1)\psi^\dagger(y_2)\pi_-^\dagger(y_3)|0\rangle_c\, ,
\en
where the CM and relative momenta of nucleon pairs are again given by
Eq.~(\ref{eq:CM}), and $p_3,q_3$ denote the pion momenta.
Near the mass shell $P^0\to M_d+{\bf P}^2/2M_d$,
$Q^0\to M_d+{\bf Q}^2/2M_d$, the six-point function (\ref{eq:6P}) develops
the double deuteron pole.
Since we are interested only in the threshold
scattering amplitude, we may take from the beginning ${\bf P}={\bf Q}=0$ and
${\bf p}_3={\bf q}_3=0$.
Then, in the vicinity of the pole, one has
\eq\label{eq:6P_pole}
G(P,Q;p,q;p_3,q_3)
&\rightarrow&\sum_{i,j}\frac{i\Psi_i(p)}{M_d-P^0}\,
G_{ij}\,
\frac{i\Psi_j^\dagger(q)}{M_d-Q^0}
\nonumber\\[2mm]
&+&\mbox{regular terms}\, ,
\en
with
\eq\label{eq:Gij}
&&(2\pi)^D\delta^D(P+p_3-Q-q_3)\, G_{ij}
=i^2\int d^Dx_3d^Dy_3
\nonumber\\[2mm]
&&\times\,{\rm e}^{ip_3x_3-iq_3y_3}
\langle B,i|T\pi_-(x_3)\pi_-^\dagger(y_3)|B,j\rangle\, .
\en
The residue of the quantity $G_{ij}$ on the pion mass shell yields the
threshold $\pi d$ scattering amplitude
\eq\label{eq:lim}
\hspace*{-.5cm}\lim_{p_3^0,q_3^0\to M_\pi}2M_\pi(M_\pi-p_3^0)(M_\pi-q_3^0)G_{ij}
=i\delta_{ij}\,{\cal T}_{\pi d}\, .
\en
On the other hand, in the theory with the Lagrangian Eq.~(\ref{eq:L_HP})
the 6-point Green function for vanishing 3-mo\-men\-ta 
can be given in the following form
\eq\label{eq:tilde}
&&G(P,Q;p,q;p_3,q_3)=\frac{2P_i^\dagger}{(m-p_1^0)(m-p_2^0)(M_\pi-p_3^0)}
\nonumber\\[2mm]
&&\times\,\frac{i(2\pi)^D\delta^D(P+p_3-Q-q_3)
R_{ij}(P,Q;p_3,q_3)}{(1-c_1J(M_d,{\bf 0}))^2}
\nonumber\\[2mm]
&&\times\,\frac{2P_j}{(m-q_1^0)(m-q_2^0)(M_\pi-q_3^0)}
+G_1\, .
\en

\begin{figure}[t]
\begin{center}
\includegraphics[width=7.5cm]{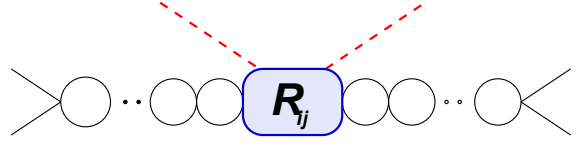}
\end{center}
\caption{Definition of the ``truncated'' Green function $R_{ij}$, 
Eq.~(\ref{eq:tilde}). The solid and the dashed lines denote nucleons and pions,
respectively.}
\label{fig:Rtilde}
\end{figure}

\noindent
The quantity $R$
in Eq.~(\ref{eq:tilde}) corresponds to the ``truncated'' Green function,
and the factor $(1-c_1J(M_d,{\bf 0}))^{-2}$ emerges after the resummation
of the $NN$ bubbles in the $^3S_1$ state before the first and after the 
last interaction of the pion with one of the nucleons 
(see Fig.~\ref{fig:Rtilde}).
Finally, $G_1$ denotes the sum of all diagrams in which the virtual 
scattering of 
the pion on one of the nucleons occurs before (or after) all $NN$
interactions, or in which the first (or last) $NN$ interaction 
happens in the $^1S_0$ state (the corresponding vertex is proportional to 
$c_0$). This class of the diagrams does not develop a double deuteron
pole, and contributes only to the regular part of the Green function. 
Consequently, from the comparison of Eq.~(\ref{eq:tilde})
to Eqs. (\ref{eq:6P_pole}), (\ref{eq:Gij}) and (\ref{eq:lim}) one may
read of the scattering amplitude at threshold
\eq\label{eq:T_pid}
\hspace*{-.4cm}\delta_{ij}{\cal T}_{\pi d}=
{\cal N}R_{ij}(P,Q;p_3,q_3)\biggr|_{P^0=Q^0=M_d,~p_3^0=q_3^0=M_\pi}\, ,
\en
where
\eq\label{eq:N}
{\cal N}=c_1^{-2}2M_\pi Z=
\frac{\gamma^d}{(4\pi)^{\frac{d}{2}}}\,\,
\frac{\Gamma(1-d/2)}{1-d/2}\,\,2M_\pi\, .
\en
Hence, the prescription for calculating the threshold
$\pi d$ scattering amplitude is formulated as follows: in the connected
6-point Green function Eq.~(\ref{eq:6P}) omit all Feynman diagrams, where
the very first or very last interaction occurs between the pion and nucleon,
or between $NN$-pair in the $^1S_0$ state. In the remaining diagrams, 
resum all initial-
and final-state $NN$ bubbles and write the final result in a form
of Eq.~(\ref{eq:tilde}); read off the quantity $R_{ij}(P,Q;p_3,q_3)$;
perform the mass-shell limit, let all 3-momenta vanish, multiply
by the normalization factor ${\cal N}$, given by Eq.~(\ref{eq:N}) and
get the threshold scattering amplitude ${\cal T}_{\pi d}$.

\subsection{Leading order}

\begin{figure}[t]
\begin{center}
\includegraphics[width=3.cm]{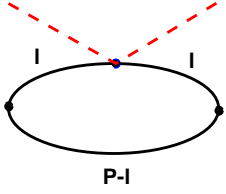}
\end{center}
\caption{Leading-order contribution to the $\pi d$ scattering length.}
\label{fig:W}
\end{figure}

At the lowest order in the expansion parameter $x=\gamma/M_\pi$, 
there is a single contribution to the quantity
$R_{ij}$ defined by Eq.~(\ref{eq:tilde}), which is depicted in 
Fig.~\ref{fig:W}.
At threshold, this contribution equals to
\eq\label{eq:W}
&&R^{(0)}_{ij}\biggr|_{\rm thr}
=4c_1^2d_+{\rm Tr}(P^\dagger_iP_j)\int\frac{d^Dl}{(2\pi)^Di}\,
\frac{1}{(w({\bf l})-l^0)^2}
\nonumber\\[2mm]
&&\times\,\frac{1}{w({\bf l})-M_d+l^0}
=\frac{c_1^2d_+m^{\frac{3}{2}}\epsilon^{-\frac{1}{2}}}{4\pi}\,\delta_{ij}\, .
\en
Substituting this result into Eq.~(\ref{eq:T_pid}), using 
Eqs.~(\ref{eq:scl}), (\ref{eq:apm}) and (\ref{eq:N}) and
expanding $M_d=2m+O(\epsilon)$, at the leading order we finally obtain
\eq\label{eq:Weinberg}
a^{(0)}_{\pi d}=\frac{1+M_\pi/m}{1+M_\pi/2m}\, 2a_++O(x)\, ,
\en
which of course coincides with the well-known result. Here we only wish to 
note that our result is valid at all orders in the chiral expansion for the
scattering length $a_+$. On the other hand, if one works in the Weinberg 
scheme, one has to identify the contributions to the quantity $a_+$ order
by order in the chiral expansion \cite{Bernard}.

\subsection{Next-to-leading order}

\begin{figure}[t]
\begin{center}
\includegraphics[width=8.cm]{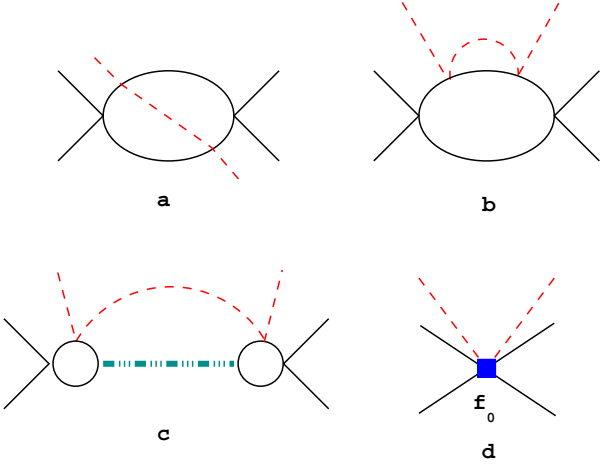}
\end{center}
\caption{Next-to-leading order contributions to the $\pi d$ scattering length, 
see Eq.~(\ref{eq:NLO}):
(a) the standard double-scattering contribution, (b) rescattering on a 
single nucleon, (c) pion rescattering 
on the infinite chain of the nucleon bubbles (``virtual deuteron'', 
denoted by a thick dot-dashed line), 
(d) counterterm contribution.}
\label{fig:NLO}
\end{figure}

Since the quantity $a_+$ turns out to be very small, the corrections
exceed in magnitude the leading-order result. In the HP EFT, 
four diagrams depicted in Fig.~\ref{fig:NLO} contribute at next-to-leading 
order. At threshold, we get
\eq\label{eq:NLO}
R^{(1)}_{ij}&\!=\!&\delta_{ij}c_1^2(R_a+R_b+R_c+R_d)\, ,
\nonumber\\[2mm]
R_a&\!=\!&2(d_+^2-2d_-^2)\int\frac{d^Dl}{(2\pi)^Di}\,\frac{d^Dk}{(2\pi)^Di}\,\,\,
\frac{1}{w({\bf l})-l^0}
\nonumber\\[2mm]
&\!\times\!&\frac{1}{(w({\bf l})-M_d+l^0)
(l^0-k^0-M_d+({\bf l}-{\bf k})^2/2M_\pi)}
\nonumber\\[2mm]
&\!\times\!&\frac{1}{(w({\bf k})+k^0)(w({\bf k})-M_d-k^0)}\, ,
\nonumber\\[2mm]
R_b&\!=\!&2(d_+^2+2d_-^2)\int\frac{d^Dl}{(2\pi)^Di}\,\frac{d^Dk}{(2\pi)^Di}\,\,\,
\frac{1}{w({\bf l})-l^0}
\nonumber\\[2mm]
&\!\times\!&\frac{1}{(w({\bf l})-M_d+l^0)^2
(w({\bf l}-{\bf k})-M_d-M_\pi+l^0-k^0)}
\nonumber\\[2mm]
&\!\times\!&\frac{1}{M_\pi+k^0+{\bf k}^2/2M_\pi}\, ,
\nonumber\\[2mm]
R_c&\!=\!&4d_+^2\int\frac{d^Dl}{(2\pi)^Di}\,\frac{d^Dk}{(2\pi)^Di}\,
\frac{d^Dq}{(2\pi)^Di}\,\,\,
\frac{1}{w({\bf l})-l^0}
\nonumber\\[2mm]
&\!\times\!&\frac{1}{(w({\bf l})-M_d+l^0)
(w({\bf l}-{\bf q})-M_d-M_\pi+l^0-q^0)}
\nonumber\\[2mm]
&\!\times\!&\frac{c_1}{1-c_1J(M_d+M_\pi+q^0,-{\bf q})}
\,\frac{1}{M_\pi+q^0+{\bf q}^2/2M_\pi}
\nonumber\\[2mm]
&\!\times\!&\frac{1}{w({\bf k}-{\bf q})-M_d-M_\pi+k^0-q^0}
\nonumber\\[2mm]
&\!\times\!&
\frac{1}{(w({\bf k})-k^0)(w({\bf k})-M_d+k^0)}
\nonumber\\[2mm]
R_d&\!=\!&\frac{1}{c_1^2}\,f_0\, .
\en
Note that performing (formally) the limit $M_\pi/m\to 0$ in the quantity
$R_a$, one gets $1/(l^0-k^0-M_d+({\bf l}-{\bf k})^2/2M_\pi)\rightarrow
2M_\pi/({\bf l}-{\bf k})^2$. In this limit, it is possible to relate the
quantity $R_a$ to the average of the operator $1/{\bf q}^2$ between the 
deuteron wave functions in the potential that corresponds to  pointlike
interaction $c_1(\psi^TP_i\psi)^\dagger(\psi^TP_i\psi)$.
In the same normalization as in 
Ref.~\cite{Bernard} one obtains
\eq\label{eq:Ra}
R_a\rightarrow 2(d_+^2-2d_-^2)\frac{2M_\pi}{Z}\,\,\frac{1}{(2\pi)^3}\,
\left<\frac{1}{{\bf q}^2}\right>_{\rm w.f.}\, ,
\en
and the standard expression for the double-scattering 
contribution in the limit $M_\pi/m\to 0$ (see e.g. \cite{Bernard})
is reproduced.

\begin{sloppypar}
The counting of the above diagrams proceeds as follows.
According to Eq.~(\ref{eq:pole}), the coupling $c_1$ counts like 
$x^{-1}$, and the couplings $d_\pm$ count like 
$x^0$. Further, after integrating over the time-like components
$l^0,k^0,q^0$, one may rescale ${\bf l}\to\gamma{\bf l}$, 
${\bf k}\to\gamma{\bf k}$, ${\bf q}\to\gamma{\bf q}$, with 
$\gamma=O(x)$. Each propagator of a pion or a nucleon
counts as $\gamma^{-2}\sim x^{-2}$ and the ``virtual deuteron
propagator'' $c_1/(1-c_1J)$ counts as $c_1\sim x^{-1}$. 
With these counting rules, it is straightforward to ensure that
$R_a\sim R_b\sim R_c=O(x^0)$ (modulo logarithms). Furthermore, since the constant $f_0$
cancels the ultraviolet divergences in the diagrams \ref{fig:NLO}a,b,c,
it must count at the same order in $x$. This fixes
$f_0=O(x^{-2})$.  Note also that
the contributions proportional to the coupling constant $c_0$ ($nn\pi^0$ 
intermediate states) have been dropped
from $R_c$ altogether at this order. This is related to the orthogonality 
of the projectors $P_i$ and $P_a$ given in Eq.~(\ref{eq:Projectors}).
\end{sloppypar}

\begin{figure}[t]
\begin{center}
\includegraphics[width=6.8cm,angle=-90]{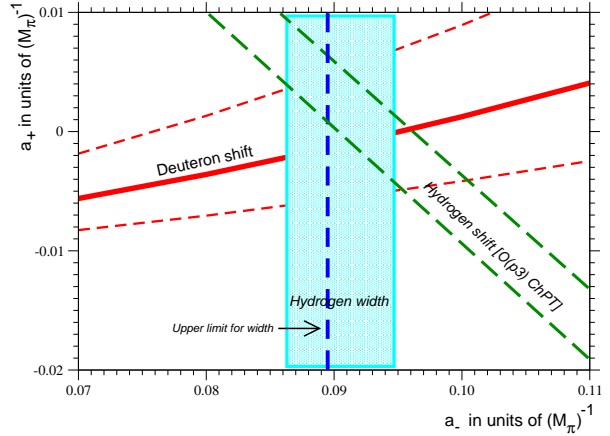}
\end{center}
\caption{The constraints on the $S$-wave $\pi N$ scattering
lengths $a_+$ and $a_-$ from the pionic deuterium energy shift
for different values of the scale parameter $\mu$. The central
line is given for $\mu=146~{\rm MeV}$. The upper and lower
bounds correspond to $\mu=250~{\rm MeV}$ and
$\mu=100~{\rm MeV}$, respectively. In addition,
we display the constraints from the most recent measurement of the pionic
hydrogen shift
 and width \cite{Gotta}. The upper limit for the width is measured
in the $4p-1s$ transition
 in pionic hydrogen.
}
\label{fig:plot}
\end{figure}

The fact that the quantity $R_c$ in Eq.~(\ref{eq:NLO}) is proportional to
$a_+^2$ simplifies the calculations considerably. Since $a_+$ is very small,
we shall systematically neglect $a_+^2$ in all expressions, and thus
assume $R_c=0$. Evaluating the remaining integrals in dimensional 
regularization and carrying out the renormalization in a standard manner,
we finally arrive at the following expression for the $\pi d$ scattering 
length at the next-to-leading order
\eq\label{eq:a_NLO}
a_{\pi d}^{(1)}&=&-\frac{m(1+M_\pi/m)^2}{\pi(1+M_\pi/2m)}\,
x\, a_-^2\,
\biggl[(x_a-x_b)\ln\frac{m\epsilon}{\mu^2}
\nonumber\\[2mm]
&+&(J_a-J_b)\biggr]
+\frac{M_\pi^4}{4\pi^2(1+M_\pi/2m)}\,
x^3\, f_0^r(\mu)
\nonumber\\[2mm]
&+&O(x^2)\, ,
\en
where $f_0^r(\mu)$ denotes the renormalized coupling constant
\eq\label{eq:f0}
f_0&=&\frac{64\pi^3(1+M_\pi/m)^2}{\epsilon M_\pi^2}\,a_-^2\, (x_a-x_b)\lambda+f_0^r(\mu)\, ,
\nonumber\\[2mm]
\lambda&=&\frac{\mu^{2(D-4)}}{16\pi^2}\biggl(\frac{1}{D-4}
-\Gamma'(1)-\ln 4\pi\biggr)\, ,
\en
and $\mu$ denotes the scale of dimensional regularization.
Further, $x_{a,b}$, $J_{a,b}$ denote the integrals over  Feynman parameters
which depend on the dimensionless variable $M_\pi/m$ and emerge from $R_{a,b}$.
These integrals are evaluated in appendix \ref{sec:int}. Here, we only give 
their approximate values
\eq\label{eq:xJ}
\begin{array}{l l}
x_a=-1.2569, \hspace*{2.cm}& J_a=-1.1334, \\
x_b=0.5098, & J_b= -0.3731\, .
\end{array}
\en 
At present, the numerical 
value of the counterterm $f_0^r(\mu)$ is not known.
For this reason, one has to include this unknown quantity 
completely in the theoretical error and to estimate the 
uncertainty that emerges already at next-to-leading order. Most easily, this
can be done by using the renormalization group equation for the scale
dependence of $f_0^r(\mu)$, which at this order reads
\eq\label{eq:RG}
\mu\frac{d}{d\mu}\,f_0^r(\mu)=-\frac{8\pi(1+M_\pi/m)^2}{\epsilon M_\pi^2}\,
a_-^2(x_a-x_b)\, .
\en
We use the following procedure to estimate the uncertainty. Since the
hard scale of the theory is of order of $M_\pi$, we may set $f_0^r(\mu)=0$
at a some scale $\mu\simeq M_\pi$ and study the $\mu$-dependence of the
plot in the $(a_+,a_-)$-plane, which emerges from Eq.~(\ref{eq:a_NLO})
at a given (experimental) value of $a_{\pi d}$. This plot is given 
in Fig.~\ref{fig:plot}. Varying $\mu$
in a ``reasonable'' range, we may thus visualize the error that is
caused by the scale dependence. Here, we wish to note that the scale dependence
is of course not the only possible source of the theoretical uncertainty
in general. In order
to have a reliable estimate of the error (in the case of the weak scale 
dependence) one has, in addition, to use  dimensional arguments to estimate
the size
of the LECs. However, in the case of a strong scale dependence, as in the 
example considered here, additional arguments are not needed.

The results which are displayed in Fig.~\ref{fig:plot} are in a qualitative
agreement with the findings of Refs. \cite{Borasoy,Beane}.
Note that these results are obtained at the next-to-leading order in HP EFT.
It is unlikely that taking into account higher-order terms will reduce the
uncertainty due to the scale dependence. On the contrary, from the comparison
with e.g. Eq.~(18) of Ref.~\cite{Beane}, one may conclude that numerically the
most important
corrections due to the effective range at this order amount to the 
multiplication of the loops by the deuteron wave-function 
renormalization factor $Z_d=(1-\gamma r_d)^{-1}\simeq 1.7$, where 
$r_d=1.765~{\rm fm}$ is the parameter related to the effective range 
in the $^3S_1$ channel. This effect leads to further amplification of the
ambiguity related to the scale dependence.

The interpretation of the results displayed in Fig.~\ref{fig:plot}
is unambiguous. The experimental data together with the above theoretical
analysis constrain the $S$-wave $\pi N$ scattering lengths within the band
which -- for large values of $\mu$ -- also intersects with the common area of two
other bands, obtained from the data on the pionic hydrogen energy shift 
and width. However, without having in advance 
estimated the numerical value of $f_0$,
one can not use the measured value of $a_{\pi d}$ for a precise determination of
the scattering length $a_+$, that was an original motivation for studying the
pion-deuteron system. Moreover, since the constant $f_0$ parameterizes
the short-range physics, it can be only 
fixed either by using other experimental
data (different from the $\pi d$ scattering process considered here) or
through the lattice simulations.

We wish to add that,  in order to be
consistent with power-counting, the 
magnitude of the LEC $f_0^r(\mu)$ 
is parametrically enhanced by a factor $1/x^2\simeq 10$
as compared to the dimensional estimate
$f_0^r(\mu)={\rm const}/M_\pi^5$. 
Numerically, substituting the dimensional 
estimate into Eq.~(\ref{eq:NLO}), one gets
the uncertainty $\sim 3\%$ in the quantity $a_{\pi d}$, which by far
underestimates the actual uncertainty displayed in Fig.~\ref{fig:plot}.

\subsection{Imaginary part}

In the presence of the absorptive channels, the coupling constant $f_0$ is
not real (note that in order to simplify the notations, we have up to 
here always omitted the
symbol ``Re'' in this coupling constant, as well as in $a_{\pi d}$ and
other quantities, if this does not lead to the confusion).
The imaginary part of the $\pi d$ scattering length is given by
(cf. with Eq.~(\ref{eq:NLO}))
\eq\label{eq:Ima}
{\rm Im}\,a_{\pi d}
=\frac{\mu_d}{2\pi}\,\Phi_0^2\,{\rm Im}\,f_0\, ,
\en
where $\Phi_0^2=\gamma^3/2\pi$ denotes the square of the deuteron wave
function at the origin. Further, 
the imaginary part of the constant $f_0$ is directly
related to the inelastic channels, which are ``shielded'' when constructing
the HP EFT. Since we have neglected all electromagnetic effects from the
beginning, the bulk contribution to ${\rm Im}\, f_0$ is provided by the
two-neutron intermediate state, whereas the contribution from the $\gamma nn$
intermediate state is omitted. To take this fact into account, we add the
superscript ``str'' to the pertinent scattering length. To get the imaginary
part, which also includes the effect of the  $\gamma nn$ state, one writes
${\rm Im}\, a_{\pi d}={\rm Im}\, a_{\pi d}^{\rm str}\,(1+1/R_\gamma)$, where the
experimental value of the quantity $R_\gamma=2.83$, which stands for the ratio of the
cross-sections of the $\pi^-d$ transition into the $nn$ and $\gamma nn$ final
states  (the Panofsky ratio) and is treated at face value.
At the lowest order we get
\eq\label{eq:f0_str}
{\rm Im}\,f_0^{\rm str}&=&\pi\int\frac{d^3{\bf k}_1}{(2\pi)^3}\,
\frac{d^3{\bf k}_2}{(2\pi)^3}\,
|T(pn\pi^-\to nn)|^2
\nonumber\\[2mm]
&\times&(2\pi)^3\delta^4(p_1+p_2+p_3-q_1-q_2) ,
\en
\begin{sloppypar}
\noindent
where $T(pn\pi^-\to nn)$ stands for the transition matrix element
for the process $p(p_1)+n(p_2)+\pi^-(p_3)\to n(k_1)+n(k_2)$, 
and the above integral has to be evaluated at threshold 
${\bf p}_1={\bf p}_2={\bf p}_3\to 0$. After straightforward calculation we
get
\end{sloppypar}
\eq\label{eq:T2}
&&{\rm Im}\,f_0^{\rm str}=\frac{mp^\star}{4\pi}\,
|T(pn\pi^-\to nn)|^2\, ,
\nonumber\\[2mm]
&&|{\bf k}_1|=|{\bf k}_2|=p^\star=\sqrt{mM_\pi}+\cdots\, ,
\en
where the ellipses stand for the relativistic corrections.
Now we note that the above result is compatible with the well-known relation
in terms of the $\pi^-d\to nn$ inelastic cross-section at threshold
\eq\label{eq:sigma}
&&{\rm Im}\,a_{\pi d}^{\rm str}=\frac{1}{4\pi}\,\lim_{|{\bf p}_3|\to 0}
|{\bf p}_3|\,\sigma(\pi^-d\to nn)\, ,
\nonumber\\[2mm]
&&\sigma(\pi^-d\to nn)=\frac{m\,\mu_d\, p^\star}{2\pi |{\bf p}_3|}\,
|T(\pi^-d\to nn)|^2\, ,
\en
if the transition amplitudes for the processes $\pi^-d\to nn$ and 
$pn\pi^-\to nn$ are related by
\eq\label{eq:Deser_strong}
T(\pi^-d\to nn)=\Phi_0\,T(pn\pi^-\to nn)\, .
\en
The equation (\ref{eq:Deser_strong}) is nothing but the leading-order 
Deser formula, which is obtained with the assumption that the deuteron radius
is much larger than the distances relevant for the interactions in the 
$\pi NN-NN$ system. The corrections to this formula would then emerge at
$O(x)$ relative to the leading-order result. Note also that
Eq.~(\ref{eq:Deser_strong}) agrees with the result of \cite{Wilkin}, obtained
within the potential scattering theory, that serves as a good check
for the validity of HP EFT.

\setcounter{equation}{0}
\section{Matching to ChPT and numerical analysis}
\label{sec:ChPT}

\subsection{Threshold amplitude in the HP EFT}

As it was demonstrated in the previous section, within the HP EFT
it is not possible to 
get an accurate description of the $\pi d$ scattering length in terms of
the $S$-wave $\pi N$ scattering length only. The relation between these 
quantities contains a large unknown short-range contribution (three-body force
in the language of the potential scattering theory), which is parameterized
through the LEC $f_0$. 

In this section we shall address the issue whether it is possible to achieve
an increased accuracy if one treats the same problem within the Weinberg 
approach \cite{Weinberg,Bernard}. 
In the HP EFT (which is the effective theory of ChPT in the Weinberg picture
for the momenta $p\ll M_\pi$), the LECs (including $f_0$)
receive contributions from two different momentum
regions: $M_\pi<p<\Lambda$ and $p>\Lambda$
(here $\Lambda\simeq 4\pi F_\pi$ denotes the cutoff mass used in the 
Weinberg formulation, which is of order of the hard scale in ChPT). 
In the Weinberg framework, 
one has ``resolved'' the momenta at the scales $M_\pi<p<\Lambda$.
The unknown dynamics at the momenta $p>\Lambda$ is parameterized by new LECs
which are now defined at the scale $\Lambda$ instead of $M_\pi$. On dimensional
grounds, the natural size of these new LECs must be much smaller than the size
of old LECs, since $M_\pi\ll \Lambda$. Stated differently, if we start from the
LECs in the theory at a scale $\Lambda$ and calculate the LECs of the 
HP EFT in the limit of a large $M_\pi$, we must see that
the LECs of the HP EFT must be enhanced by the pion loops where the
loop momentum runs within $M_\pi<p<\Lambda$. It is natural to assume that
this momentum region contributes the bulk of the total magnitude of the LEC
in question. Could  one then separate the 
large but potentially calculable pion exchange contribution to $f_0$ from
unknown short-range contribution at a scale $\Lambda$?

A natural choice of the $S$-matrix element in the scattering sector, 
which can be used
to determine the constant $f_0$ through the matching procedure
to the Weinberg framework, is that of
the elastic process $\pi^-(pn)_{^3S_1}\to\pi^-(pn)_{^3S_1}$. 
The matching condition has the form
\eq\label{eq:match_start}
&&T_W(p_1p_2p_3;q_1q_2q_3)=\prod_{i=1}^3
[2E_i({\bf p}_i)]^{1/2}[2E_i({\bf q}_i)]^{1/2}
\nonumber\\[2mm]
&&\hspace*{.4cm}\times\,T(p_1p_2p_3;q_1q_2q_3)\, ,
\en
where $T$ and $T_W$ denote pertinent scattering matrix elements
in the HP EFT and in ChPT, respectively. Further, 
$E_i({\bf l})=\sqrt{m^2+{\bf l}^2},~i=1,2$ and
$E_3({\bf l})=\sqrt{M_\pi^2+{\bf l}^2}$ 
are the relativistic energies of a nucleon and a pion. Note that 
the kinematical factor in 
Eq.~(\ref{eq:match_start}) is introduced in order to take into account the 
different normalization of the one-particle states in HP EFT and in the
Weinberg approach. 

We found it convenient to perform matching
for the one-particle irreducible (1PI) matrix elements separately in
two- and three-particle subsystems
(with respect to the {\em non-relativistic} pion and nucleon propagators).
In the three-particle sector, the scattering amplitude turns out to be
singular at
threshold: one has to choose the particular kinematics in order to approach
the the zero-momentum limit. A possible
choice of the external momenta is given by
\eq\label{eq:kin}
{\bf p}={\bf q}=0 ,\,\, {\bf p_3}=-{\bf P} ,\,\,
{\bf q_3}=-{\bf Q} ,\,\, {\bf P}=-{\bf Q}\to 0 ,
\en
where the definition of the CM and the relative momenta is given in 
Eq.~(\ref{eq:CM}).

Below, we schematically 
describe the matching for the above scattering matrix element.
Explicit expressions will not be used in the discussion.
For our purposes, we only need to demonstrate 
that such a matching can be performed in principle.

\begin{figure}[t]
\begin{center}
\includegraphics[width=8cm]{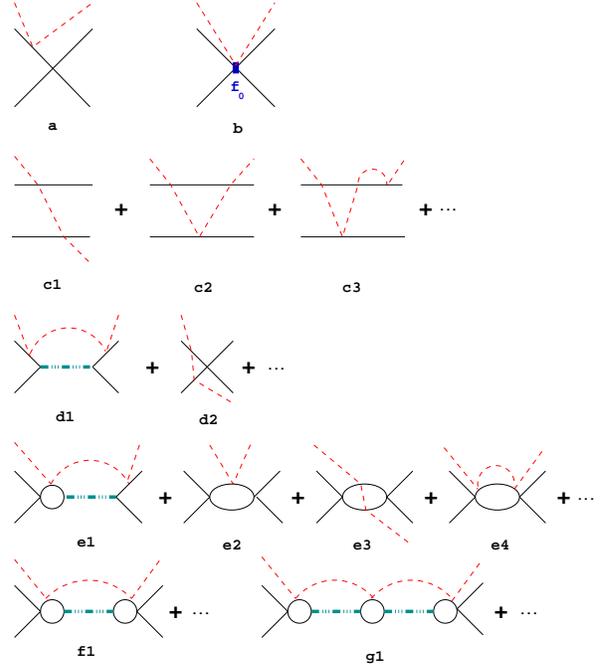}
\end{center}
\caption{Lowest-order diagrams contributing to the threshold scattering
  amplitude for the process $\pi^-(pn)_{^3S_1}\to\pi^-(pn)_{^3S_1}$. The
  solid, dashed and dot-dashed lines correspond to nucleons, to 
 pions and to the deuteron (an infinite sum of nucleon bubbles), 
respectively. All diagrams except (a) and (c1) are one-particle irreducible.}
\label{fig:ampl}
\end{figure}

We start with the evaluation of the scattering amplitude $T$ that enters the 
right-hand side of Eq.~(\ref{eq:match_start}). This quantity
receives many contributions (some of them are depicted in Fig.~\ref{fig:ampl}).
Out of these contributions, the diagrams
\ref{fig:ampl}a and \ref{fig:ampl}c1 are one-particle reducible and will be
excluded from matching.
Some of the contributions are singular at threshold.
For example, the diagram in Fig.~\ref{fig:ampl}e2 contains the factor
$|{\bf P}|^{-1}$, and the diagrams in Fig.~\ref{fig:ampl}e3,e4
are logarithmically singular at small momenta. The three-particle
threshold
scattering amplitude ${\cal A}(\mu)$, 
by definition, is obtained from the 1PI matrix element, 
by subtracting first all singular contributions at $|{\bf P}|\to 0$. Namely,
in the vicinity of threshold, the sum of the 1PI diagrams shown in 
Fig.~\ref{fig:ampl} has the form
\eq\label{eq:threshold}
T^{\rm (1PI)}=\frac{T_{-1}}{|{\bf P}|}
+T_0\ln\frac{|{\bf P}|}{\mu}+{\cal A}(\mu)+O(|{\bf P}|)\, ,
\en
where the arbitrary scale that enters the logarithm is  
set equal to the scale of dimensional regularization $\mu$, for simplicity.
Further, the quantity ${\cal A}(\mu)=f_0^r(\mu)+\overline{{\cal A}(\mu)}$ 
contains additive contribution from the LEC $f_0^r(\mu)$, 
which emerges in the diagram Fig.~\ref{fig:ampl}b. 
Note that the corrections to this contribution due to the initial- and
finite-state interactions vanish in HP EFT. Here, $\overline{{\cal A}(\mu)}$
stands for a sum of all diagrams that do not contain $f_0^r(\mu)$,
see Fig.~\ref{fig:ampl}. At the order we are working, this quantity depends
only on the couplings in the two-particle sector $d_\pm$, $c_1$, as well as
the masses $M_\pi$, $m$.

The power counting in the expansion parameter $x$ is organized as follows.
One groups the diagrams shown in Fig.~\ref{fig:ampl}, 
according to the number ($d_N$) of virtual $NN$ interactions -- 
``squeezing''  all deuteron propagators to a single point.
For example, the 1PI diagrams (c2),(c3) correspond to $d_N=0$,
(d1), (d2) -- to $d_N=1$, (e1), (e2), (e3), (e4) -- to $d_N=2$,
(f1) -- to $d_N=3$, (g1) -- to $d_N=4$, etc. 

For illustration,
let us first consider the 1PI diagrams with $d_N=0$. The contribution
to the threshold
amplitude, which is obtained after subtracting all singular pieces from these
diagrams, is proportional to powers of the $\pi N$ coupling constants
$d_\pm$ and depends on the masses $M_\pi$, $m$ and the scale $\mu$. Neither
of these quantities scale with $x$ and thus this contribution emerges at
$O(1)$.

Next, we consider 
the diagrams with $d_N=1$. The diagram (d1) is proportional to
$c_1d_\pm^2=O(x^{-1})$. In addition, there is an intrinsic
scale $x$ present in the deuteron propagator
$(1-c_1J)^{-1}$, since $c_1=O(x^{-1})$. In order to get rid
of this scale, one has to first perform the contour integration over 
the time-component of the loop momentum $k^0$
and afterwards rescale the loop three-momentum ${\bf k}\to 
x{\bf k}$, as well as all external three-momenta. Further, after
subtracting all singular pieces at threshold, one may put the external momenta to
zero. The counting of the powers of $x$ proceeds as follows. The
integration measure $d^3{\bf k}$ yields the factor $x^3$. Since the 
integration over $k^0$ reduces the total number of elementary (pion and
nucleon) propagators from three to two and since each elementary propagator
counts like $x^{-2}$, this leads to the factor $x^{-4}$ after
rescaling the momenta. Finally, taking into account the fact that after
rescaling the deuteron propagator counts like $O(1)$ and putting together all
factors, we arrive at the conclusion that the diagram (d1) counts like
$x^{-2}$ (modulo logarithms). By using the same method, it is easy to show 
that the diagram (d2) contributes at
$O(x^{-1})$.

Applying the same argument to all diagrams in Fig.~\ref{fig:ampl}, we finally
come to the conclusion that the leading order scaling of the loop diagrams
is that at $O(x^{-2})$. 
Only the diagrams (d1), (e1), (e3), (e4), (f1) contribute at the leading
order in $x$. All other diagrams, 
as well as the diagrams which can be obtained from the
diagrams displayed in Fig.~\ref{fig:ampl} by attaching more pion and/or nucleon
loops, contribute at a higher order in $x$.

Finally one has to consider the LEC $f_0^r(\mu)$ which, as we know,
scales at $O(x^{-2})$ at leading order. Generally, one can write
\eq\label{eq:f0_eps}
f_0^r(\mu)&=&F_{-2}x^{-2}+F_{-1}x^{-1}+F_0+F_1x+\cdots\, ,
\nonumber\\[2mm]
F_i&=&F_i(M_\pi,m,\mu)\, ,
\en
where only the leading coefficient $F_{-2}$ contributes in the pion-deuteron 
scattering length at $O(x)$.

To summarize, 
the right-hand-side of Eq.~(\ref{eq:match_start}) can be written as
a sum of the terms that scale as $x^n$ with $n=-2,-1,0,\cdots$.  
The leading-order contribution at $O(x^{-2})$ emerges from the diagrams 
(b), (d1), (e1), (e3), (e4) and (f1). Except the LEC $f_0$ that enters
from diagram (b), all other diagrams are expressed in terms of the parameters
determined in the two-particle sector.

\subsection{Weinberg-Tomozawa term}

\begin{figure}[t]
\begin{center}
\includegraphics[width=2.5cm]{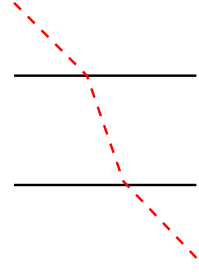}
\end{center}
\caption{Lowest-order contribution to the threshold amplitude of the process
$\pi^-pn\to\pi^-pn$ in ChPT. The solid and dashed lines
denote nucleons and  pions, respectively.}
\label{fig:Weinberg}
\end{figure}

Next, we consider the evaluation of the left-hand side of 
Eq.~(\ref{eq:match_start}). The lowest-order contribution in ChPT in the 
Weinberg framework emerges from the diagram
depicted in Fig.~\ref{fig:Weinberg}. This diagram describes the
double scattering of the pion on the nucleons,
with the vertices obtained from Weinberg-Tomozawa Lagrangian
\eq\label{eq:LWT}
{\cal L}&=&\frac{i}{2\sqrt{2}F^2}\,\bigl\{
 (\pi^+\partial_\mu\pi^0-\pi^0\partial_\mu\pi^+)\bar p\gamma^\mu n
\nonumber\\[2mm]
&+&(\pi^-\partial_\mu\pi^0-\pi^0\partial_\mu\pi^-)\bar n\gamma^\mu p\bigr\}
\nonumber\\[2mm]
&+&\frac{i}{4F^2}\,(\pi^-\partial_\mu\pi^+-\pi^+\partial_\mu\pi^-)
\bigl\{\bar p\gamma^\mu p-\bar n\gamma^\mu n\bigr\}\, ,
\en
where one has used the Condon-Shortley convention for the component fields and
$F$ stands for the pion decay constant in the chiral limit. At the order of
accuracy we are working, we may take $F=F_\pi=92.4~{\rm MeV}$.
The scattering amplitude for the process 
$p(q_1)n(q_2)\pi^-(q_3)\to p(p_1)n(p_2)\pi^-(p_3)$ at second
order in perturbation theory is given by
\eq\label{eq:WT}
T_{W}&=&-\,\frac{1}{4F^4}\,\frac{\bar u(p_1)\!\not\! q_3 u(q_1)\,
\bar u(p_2)\! \not\! p_3 u(q_2)}{M_\pi^2-(p_1-q_1-q_3)^2}
\nonumber\\[2mm]
&-&\frac{1}{4F^4}\,\frac{\bar u(p_1)\!\not\! p_3 u(q_1)\,
\bar u(p_2)\! \not\! q_3 u(q_2)}{M_\pi^2-(p_1-q_1+p_3)^2}
\nonumber\\[2mm]
&-&\frac{1}{2F^4}\,\frac{\bar u(p_1)\!\not\! p_3 u(q_2)\,
\bar u(p_2)\! \not\! q_3 u(q_1)}{M_\pi^2-(p_1-q_2+p_3)^2}\, .
\en

The low-energy reduction of the relativistic
amplitude Eq.~(\ref{eq:WT}) yields both the reducible and irreducible parts.
In order to single out the 1PI piece, one has to split the relativistic pion 
propagator into the positive- and negative-energy components and expand the
result for small three-momenta. For example,
\eq\label{eq:splitting}
&&\frac{1}{M_\pi^2-(p_1-q_1-q_3)^2}
\nonumber\\[2mm]
&\!=\!&\frac{1}{2M_\pi}\,
\frac{1}{M_\pi-q_1^0-q_3^0+p_1^0+({\bf q}_1+{\bf q}_3-{\bf p}_1)^2/2M_\pi}
+\cdots
\nonumber\\[2mm]
&\!+\!&\frac{1}{4M_\pi^2}+\cdots\, ,
\en
where the first (the second) term correspond to the one-particle reducible
(irreducible) pieces. Evaluating the 1PI part at threshold, we get
\eq\label{eq:1PI_W}
T_{W}^{\rm (1PI)}=-\frac{m^2}{2F^4}\,(\chi_1^\dagger\chi_1\,\chi_2^\dagger\chi_2
+\chi_1^\dagger\chi_2\,\chi_2^\dagger\chi_1)\, ,
\en
where the Pauli spinors are defined through 
$\sqrt{2m}\begin{pmatrix}{\chi_i\cr 0}\end{pmatrix}=
\lim_{q_i\to 0}u(q_i)$ and 
$\sqrt{2m}\,\,(\chi_i^\dagger,0)=\lim_{p_i\to 0}\bar u(p_i)$. 
Performing a Fierz transformation,
we get
\eq\label{eq:1PI_Theta}
T_{W}^{\rm (1PI)}=-\frac{m^2}{F^4}\,(\chi P_i \chi)^\dagger
(\chi P_i \chi)\, ,\quad\quad 
\chi=\begin{pmatrix}{\chi_1\cr \chi_2}\end{pmatrix}\, .
\en
Finally, dividing the amplitude $T_W$ by a kinematical 
factor $(2m)^22M_\pi$ that emer\-ges from Eq.~(\ref{eq:match_start}) at 
threshold, one may read off the value of the LEC $f_0$, which is obtained
through the matching to ChPT at the leading order in chiral expansion
\eq\label{eq:f0_ChPT}
f_0=-\frac{1}{8F^4M_\pi}+\cdots\, .
\en
Substituting now this value into Eq.~(\ref{eq:NLO}), one
ends up with the tiny contribution $-0.0005~M_\pi^{-1}$ (cf. with the
current experimental value $a_{\pi d}=0.0261~M_\pi^{-1}$). 
The small magnitude for this correction is in agreement with 
findings of Ref.~\cite{Weinberg}.
The reason why the leading-order chiral contribution is so small is simple:
it emerges only at the NNLO in the $x$ counting and is contained
in the term $F_0$ of Eq.~(\ref{eq:f0_eps}). To the contrary, some of the
sub-leading contributions in ChPT, which are contained in the LO and NLO
coefficients $F_{-2}$, $F_{-1}$, get enhanced by inverse powers of $x$ and
are numerically much larger that the Weinberg-Tomozawa term. The above example
clearly shows that the power-counting in the HP EFT and in the Weinberg picture
are not correlated. In accordance with the findings of Ref.~\cite{Bernard},
the counting which is based on the expansion parameter $x$, better reflects
the numerical size of the contributions emerging at different orders.

\subsection{Initial- and final-state interactions}

\label{sec:IF}

From the discussion in the previous section one concludes that one needs
to identify a sub-class of the diagrams in the Weinberg approach which
-- after the matching -- contributes at $O(x^{-2})$ to the LEC $f_0$.
Since here we are only interested in establishing accuracy limits in the
calculations of the pion-deuteron scattering length, in the matching condition 
we may neglect loop diagrams shown in Fig.~\ref{fig:ampl} and their 
counterparts in ChPT. Since these diagrams depend only on the parameters that
can be determined in the two-particle sector, including these diagrams will
only shift the central value of $f_0$ determined through matching, without
significantly affecting the error bars.

The contributions which one needs to retain in the left-hand side
of the matching condition Eq.~(\ref{eq:match_start}), contain the genuine
3-particle ChPT LEC(s) $f_0'$, which are the counterpart(s) of the LEC $f_0$
in HP EFT. Note that due to chiral symmetry, the LECs in the Weinberg 
framework appear first at $O(p^2)$ -- the corresponding Lagrangian should 
contain either two derivatives of the pion field or the quark mass matrix.
Further, these LECs do not come alone: in the spirit of the
Weinberg approach, one has in addition to consider summing up the diagrams
that correspond to the
 initial- and final- state interactions of pions and nucleons.
 Due to the use of the cutoff in the Weinberg approach, 
this effect does not vanish at threshold, in difference to HP EFT.

\begin{figure}[t]
\begin{center}
\includegraphics[width=7.5cm]{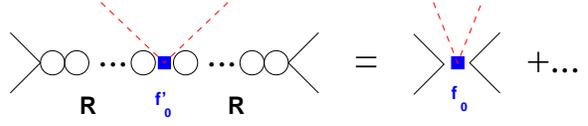}
\end{center}
\caption{Including the initial- and final-state interactions in the 
Weinberg picture}
\label{fig:bubbles}
\end{figure}

In the following, we shall first provide a very crude estimate
of the initial- and final-state interaction effect. We assume
that the bulk of the effect in the 3-particle system comes from the $NN$
pair, whereas the ``hopping'' of the pion on the nucleon lines amounts 
to a small correction. We further approximate the one-pion exchange in the Weinberg picture by a local $NN$ potential with a coupling denoted by $c_1'$, 
assuming a cutoff between the pion mass $M_\pi$ and the hard scale $\Lambda$
in the loops, wherever this vertex is present. With these assumptions,
the matching condition takes the form that is schematically shown in 
Fig.~\ref{fig:bubbles} (in this figure, the ellipses stand for the diagrams
which do not contain 3-particle LECs). One may introduce the ``amplification
factor'' which is obtained by summing up all bubbles in the ingoing and
outgoing $NN$ lines
\eq\label{eq:R}
R^2&=&\frac{1}{(1-c_1'(\Lambda)J_\Lambda(0))^2}\, ,
\nonumber\\[2mm]
J_\Lambda(|{\bf p}|)&=&\int^\Lambda\frac{d^3{\bf l}}{(2\pi)^3}\,
\frac{m}{{\bf l}^2-{\bf p}^2-i0}
\nonumber\\[2mm]
&=&\frac{m\Lambda}{2\pi^2}+i\,\frac{m|{\bf p}|}{4\pi}
+O(\Lambda^{-1})\, ,
\en
where we have made explicit the $\Lambda$-dependence of the coupling constant
in the Weinberg scheme.

Imposing now the condition that the denominator has a
pole at the deuteron binding energy $|{\bf p}|=i\gamma$,  one can
determine the coupling constant $c_1'(\Lambda)$ and the amplification factor
\eq\label{eq:c1Lambda}
&&(c_1'(\Lambda))^{-1}=\frac{m\Lambda}{2\pi^2}-\frac{m\gamma}{4\pi}+\cdots\, ,
\nonumber\\[2mm]
&&R^2=\biggl(1-\frac{2\Lambda}{\pi\gamma}\biggr)^2+\cdots\approx
\frac{4\Lambda^2}{\pi^2 m\epsilon}+\cdots\, ,
\en
and the matching gives 
\eq\label{eq:f0_x2}
f_0\propto R^2f_0'+\cdots=r^2x^{-2}f_0'+\cdots\, ,
\en 
where $r^2$ does not scale with $x$. Hence, the desired scale behavior in
the constant $f_0$ emerges from the sub-class of the diagrams in the 
Weinberg picture which describe $NN$ rescattering in the initial and final
states in all orders.

The numerical value of the amplification factor $R^2$ turns out to be very 
large: $R^2\simeq 4~{\rm for}~\Lambda=M_\pi$,
$R^2\simeq 50~{\rm for}~\Lambda=500~{\rm MeV}$ and
$R^2\simeq 200~{\rm for}~\Lambda=1~{\rm GeV}$.
Performing now a standard dimensional analysis of the LECs in ChPT 
(see e.g. \cite{Friar}), we get the estimate of the uncertainty
in the quantity $f_0$
\eq\label{eq:Friar}
\Delta f_0= \frac{1}{2M_\pi}\, R^2\Delta f_0'=\frac{1}{2F_\pi^4 M_\pi}\,
\biggl(\frac{M_\pi}{4\pi F_\pi}\biggr)^2\,R^2\, .
\en
Substituting this result into Eq.~(\ref{eq:NLO}), one finally gets
\eq\label{eq:thumb}
&&\Delta a_{\pi d}/a_{\pi d}=[0.5\%;6\%;23\%]
\nonumber\\[2mm]
&&\mbox{for}\quad
\Lambda=[M_\pi;500~{\rm MeV};1~{\rm GeV}]\, .
\en
Now we present the exact numerical results in the Weinberg picture 
which confirm the validity of our crude estimate. 
Note that, if the calculations 
are done in the Weinberg approach {\it ab initio},
the amplification factor is contained in the quantity
$|\tilde\Phi_0(\Lambda)|^2$, where $\tilde\Phi_0(\Lambda)$
denotes the value of the wave function at the origin.
The counterpart of this quantity in HP EFT is given by
$\Phi_0=(\gamma^3/2\pi)^{\frac{1}{2}}=
0.0445~{\rm fm}^{-\frac{3}{2}}$. 
For comparison, the NNLO wave functions in
the Weinberg approach are given by\\ 
$\tilde\Phi_0(\Lambda)
=[0.487;0.434]~{\rm fm}^{-\frac{3}{2}}$ for $\Lambda=[450;650]~{\rm MeV}$
\cite{Krebs-private}. 
Hence, one gets the following estimate for the amplification
factor $R^2=\tilde\Phi_0(\Lambda)^2/\Phi_0^2\simeq 100$, which
in turn corresponds to the reasonable value of the 
cutoff mass $\Lambda\simeq 720~{\rm MeV}$
in our order-of-magnitude estimate in Eq.~(\ref{eq:thumb}) and to
 $\Delta a_{\pi d}/a_{\pi d}\simeq 12\%$, which 
qualitatively agrees with the large uncertainty in Fig.~\ref{fig:plot}.
Note that the $NN$ interactions are now studied up to and including 
N$^3$LO~\cite{Epelbaum}. In our estimates we however use the NNLO
wave functions, in order to consistently compare with earlier 
work on the subject~\cite{Bernard}.

The LEC $f_0$ that enters the expression of the pion-deuteron scattering 
length, can be estimated by using resonance saturation. The details of this
procedure can be found in appendix~\ref{sec:resonance}.
It turns out that using the set of parameters determined in Ref.~\cite{BKM}
leads to the result which is consistent with the above dimensional
estimate. However, due to the large uncertainty in the values of these 
parameters, the resonance saturation hypothesis can not be used -- at present stage --
to get a better accuracy in the calculation of $a_{\pi d}$. To this
end, one has to improve the quality of the fit of the parameters of the 
resonance Lagrangian to the data.

How can this large uncertainty be reconciled with the mild cutoff dependence
of Ref.~\cite{Bernard}? The answer to this question is the following.
Assuming for instance 
that the cutoff dependence of the generic LEC(s) $f_0'$ in ChPT 
at the lowest order is logarithmic, we may split these LEC(s) into a term 
which depends of the cutoff mass $\Lambda$ and a term which does not
\eq\label{eq:AB}
 f_0'=A+B\ln\frac{\Lambda}{\Lambda^*}\, ,
\en
where $\Lambda^*$ stands for a some characteristic hard scale (typically 
$\sim 1~{\rm GeV}$). The mild cutoff dependence is equivalent to the statement
that the constant $B$ is small. One may argue that the 
smallness of $B$ is related
to the dominance of the one-pion exchange which at the large distances 
has a softer behavior than the contact 4-fermion vertex which is used
in HP EFT to describe the deuteron. However, the constant $A$ is not covered
by above argument. Using the dimensional arguments, we see that the bulk
of the uncertainty $\sim 12\%$ should come from the scale-independent
constant $A$.

\setcounter{equation}{0}
\section{Comparison to the existing approaches}
\label{sec:lit}

The issue of the pion-deuteron scattering has obtained an extensive coverage
in the literature during the last decades. A highly incomplete list of
references is given in 
\cite{Afnan,Faldt,Mizutani,Thomas,Deloff,Baru,Loiseau,Hanhart,Wilkin},
see also references therein. 
Note that the framework used in HP EFT is in fact very close to that of
the potential scattering theory. Consequently, a rapid convergence of the
series for $a_{\pi d}$ in HP EFT would  indicate also on the
applicability of the potential picture in the $\pi d$ scattering at threshold.
The main question related to choice of the potential
 remains however open. As was
demonstrated in Ref.~\cite{Lipartia}, the conventional quantum-mechanical
potentials can be interpreted as a mere regularization of the non-relativistic
effective field theories. From this point of view, the scale dependence
which was discovered in this article is equivalent to the off-shell
effects in the two-body potentials which must be canceled by the corresponding
three-body force (analog of the LEC $f_0$). It is clear that fitting
two-body potentials to the scattering data can not completely eliminate this
inherent off-shell (scale) dependence. Another aspect of the
problem concerns the absence of the relation to QCD and to ChPT. 
One may
conclude  that the existing potential models which are used to extract
 $\pi N$ scattering lengths from the measured $\pi d$ scattering length
can not in principle
 provide enough accuracy needed for the test of the predictions of QCD at low
 energy in the $\pi N$ sector.

Next, we shall consider the ``hybrid'' approach in nuclear physics, 
which is based on the
calculation in ChPT of a certain set of Feynman diagrams, corresponding
to the ``irreducible transition kernel'' and finally 
sandwiching the result by ``realistic'' wave functions which are calculated,
using  Paris, Bonn, Argonne$,\ldots$ potentials (see
e.g.\cite{Hanhart,Weinberg,Doring}). As we already mentioned above, this
procedure can be justified, if and only if the long-range effects 
(e.g. the one-pion exchange) dominate the transition operator. However, as we
have seen, the situation in describing $\pi d$ scattering length is just the
opposite: this quantity receives a large short-distance contribution from the
LEC $f_0$, which should be there in order to provide a scale-independence of
the final result. In the final expression for the decay width, this constant
is multiplied by the wave function of the deuteron at the origin squared.
Consequently, for the consistency of the hybrid approach, one must be sure
that the value of
the ``realistic'' wave function at the origin 
is a good approximation of the same quantity, obtained
in the effective theory which was previously used to calculate the
transition kernel.

\begin{sloppypar}
A fully consistent approach to the pion-deuteron scattering problem is
provided by the effective field theories, in which the wave function of the
deuteron is evaluated within the same setting as the diagrams, describing
the irreducible kernel. These are e.g. the calculations carried out within
the framework with perturbative pions \cite{Borasoy}, HP EFT with elementary
deuteron field \cite{Beane}, or in the Weinberg scheme \cite{BBLM,Bernard}. 
The large scale
dependence in the LECs analogous to our $f_0$, was first reported in
Ref.~\cite{Borasoy}. The Feynman diagram which leads to such a large
dependence is the counterpart of our Fig.~\ref{fig:NLO}a. One should however
note that treating the pions relativistically and using the chiral Lagrangians
unnecessarily complicates the simple physical picture, since the question
about the convergence of ChPT expansion naturally arises. 
A more straightforward approach is provided by HP EFT
\cite{Beane}. The authors of
this article also find the logarithmic enhancement of the diagram
\ref{fig:NLO}a, as well as the large scale dependence which emerges due to
this diagram. Note however, that in this paper not all terms at $O(x)$ have
taken into account: The diagrams shown in Fig.~\ref{fig:NLO}b,c are omitted --
although, as we have seen, the inclusion of these diagrams does not change
the qualitative conclusions. In the present paper, we have also critically
re-examined the conjecture made in Ref.~\cite{Beane}, concerning the
possibility of calculation of the LEC $f_0$ with an improved accuracy
in the Weinberg picture -- this, as was shown, is not possible.
\end{sloppypar}

\begin{sloppypar}
Finally, we briefly comment on the papers \cite{BBLM,Bernard}, which
provide the systematic treatment of the $\pi d$ scattering problem within the
Weinberg framework. The results of these papers 
are now very easy to understand
and the relation to the HP EFT  becomes crystal clear.
Thus, the puzzle concerning the (seeming) differences
between the HP EFT and the Weinberg approach has finally been resolved.
\end{sloppypar}

\setcounter{equation}{0}
\section{Conclusions}
\label{sec:concl}

\begin{itemize}

\item[i)]
In this paper, we systematically investigate pionic deuterium
within the framework of effective field theories. The whole treatment
naturally falls into several steps. At the first step, we discuss the
extraction of the $\pi d$ scattering length from the
$3p-1s$ transition energy and width in the pionic deuterium. 
The next-to-leading order result has been obtained for the level energies, 
which
can be used for an accurate determination of the $\pi d$ threshold scattering
amplitude from experimental measurements. Since the isospin-breaking
corrections in this amplitude are not expected to be relevant, given the
relatively large theoretical uncertainty in connecting $a_{\pi d}$ with the
$\pi N$ scattering lengths, these corrections have been neglected for the time
being. 

\item[ii)]
The main focus in the present paper is on investigating the possibility to
relate $a_{\pi d}$ to the $\pi N$ scattering lengths and on the analysis of
the systematic theoretical error in such a procedure. We give a consistent
treatment of the problem within the framework of HP EFT, where the expansion
parameter is given by the quantity $x=\gamma/M_\pi\simeq 1/3$, where 
$\gamma\simeq 45~{\rm MeV}$ is the characteristic bound-state momentum in the
deuteron. In this paper, we have evaluated contributions to the quantity
$a_{\pi d}$, up to and including $O(x)$ that corresponds to the
next-to-leading order in HP EFT.

\item[iii)]
At next-to-leading order, 
the $\pi d$ scattering length receives a contribution
from the (unknown) LEC which we denote as $f_0$. We have used the
scale-dependence of this contribution for estimating the theoretical error in
our calculations. This scale dependence is shown in Fig.~\ref{fig:plot}.
The scale must be chosen to be of order of the pion mass, but is otherwise 
arbitrary. In our opinion, the range $100~{\rm MeV}<\mu<250~{\rm MeV}$ can 
be roughly considered as a ``natural'' choice of this scale. It should be 
also taken into account that the plot in this figure, which corresponds 
to the recent experimental measurements of the pionic hydrogen decay width, 
still does not include the isospin-breaking corrections from ChPT 
\cite{Zemp,Buettiker}. As we see from Fig.~\ref{fig:plot}, the theoretical
uncertainty due to the unknown LEC $f_0$ is rather large.

\item[iv)]
In this paper, we have investigated in detail the
differences between the Weinberg approach and the HP EFT. It was shown that,
despite the very mild cutoff dependence in the Weinberg approach, the 
uncertainty due to the unknown LECs is significant and is of the same order 
of magnitude as in the HP EFT. The reason for this is that the large 
initial- and final-state $NN$ interactions lead to the amplification 
of the initially small LEC contribution. Taking into account this 
amplification, the theoretical predictions within both  approaches are
essentially the same.

\item[v)]
Our main conclusion, concerning the accuracy limits in the
extraction of the $\pi N$ 
scattering lengths from the pion-deuteron data, can be formulated as follows:
by far the largest source of uncertainty is the low-energy constant $f_0$,
which is the genuine short-distance three-body
contribution and should be either 
determined by other experiments or should be obtained by lattice simulations.
In particular, one might attempt to get at least the order-of-magnitude
estimate from the process $NN\to NN\pi\pi$
or from the pion-nucleus optical potential, in case of non-equal
proton and neutron densities, where $f_0$ should be present apart 
from the ``conventional'' terms as given in e.g. \cite{Oset}. Neither of these
methods seems easy to be applied.
But, without having 
fixed the value of $f_0$ at a sufficient precision, it is impossible
to improve the accuracy of the prediction of the pion-deuteron scattering
length.

\item[vi)]
In our opinion,
it is feasible to estimate  $f_0$ by using the resonance saturation
hypothesis (see appendix~\ref{sec:resonance}). At present time, however, the
parameters of the resonance Lagrangian are not known at a sufficient precision,
and more effort is needed to pin them down accurately from the experimental 
data.

\end{itemize}

\begin{sloppypar}
{\it Acknowledgments:} The authors would like to thank
V.~Ba\-ru,
S.~Be\-a\-ne,
B.~Bo\-ra\-soy,
J.~Gas\-ser,
D.~Got\-ta,
C.~Han\-hart,
H.~Krebs,
A.~Kud\-ryav\-tsev,
B.~Metsch, 
E.~Oset,
L.~Plat\-ter,
M.~Sa\-va\-ge 
and
A.~Si\-bir\-tsev
for interesting discussions.
\end{sloppypar}

\appendix
\renewcommand{\thesection}{\Alph{section}}
\renewcommand{\theequation}{\Alph{section}\arabic{equation}}

\setcounter{equation}{0}
\section{Calculation of the energy shift of pionic deuterium by using 
non-relativistic Lagrangians}
\label{sec:shift}

In this appendix, we shall present the calculation of the energy of 
the levels of the pionic deuterium, characterized by the quantum numbers $nlj$.
The calculations will be performed up to and including next-to-leading order
in isospin-breaking parameters $\alpha$ and $m_d-m_u$.
Since similar calculations for other hadronic bound states have been 
already considered in the literature in great detail
(see e.g. \cite{Bern1,Bern2,Bern4,Schweizer,Bern3,Zemp,Bonn1}), 
we shall not repeat
them here and display only the most important results. For convenience, 
the energy of the generic level can be split into the ``electromagnetic''
and ``strong'' parts
\eq\label{eq:a1}
E_{nlj}&=&E_n^{\rm C}+\Delta E_{nlj}^{\rm em}+\Delta E_{nlj}^{\rm str}
+o(\alpha^4,\alpha^3(m_d-m_u))\, ,
\nonumber\\[2mm]
E_n^{\rm C}&=&-\frac{1}{2n^2}\,\alpha^2\mu_d\, ,
\en
where $E_n^{\rm C}$ stands for the non-relativistic Coulomb binding energy.

\begin{figure}[t]
\begin{center}
\includegraphics[width=8.cm]{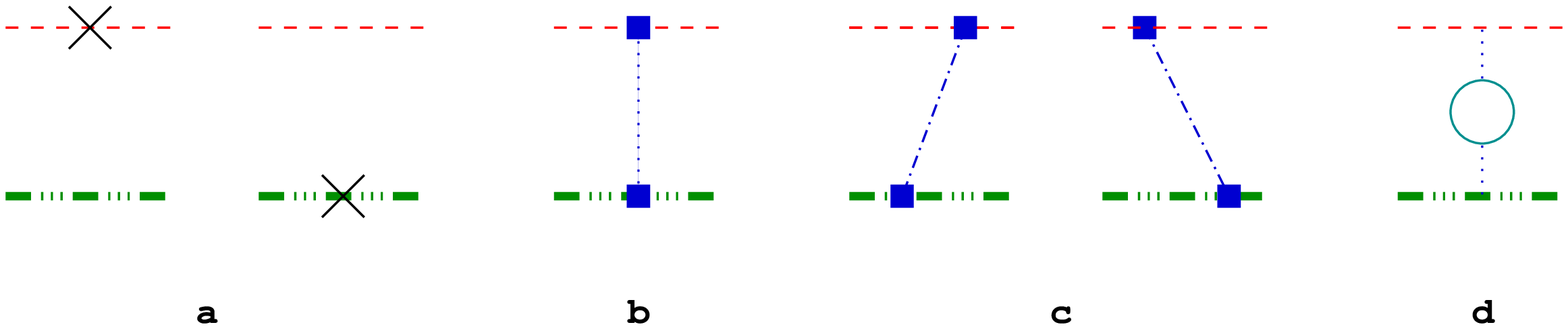}
\end{center}
\caption{The contributions to the electromagnetic energy of the pionic 
deuterium: a) relativistic insertions; b) one Coulomb photon exchange;
c) one transverse photon exchange 
(shaded squares denote pion and deuteron electromagnetic formfactors);
d) vacuum polarization.}
\label{fig:em_energy}
\end{figure}

The contributions to the ``electromagnetic'' part of the potential are depicted
in Fig.~\ref{fig:em_energy}. These include: the relativistic insertions 
${\bf p}^4/8M_\pi^3$, ${\bf p}^4/8M_d^3$ in the pion and de\-u\-te\-ron lines, 
respectively; the one (Coulomb and transverse) 
photon exchange between the pion and the deuteron;
the vacuum polarization contribution. At the end of the day, 
the explicit expression for the potential at this order can be written as
\eq\label{eq:Vem}
&&V^{\rm em}_{ab}({\bf p},{\bf q})= 
(2\pi)^3\delta^3({\bf q}-{\bf p})\delta_{ab}
\left(\frac{{\bf p}^4}{8M^3_\pi} +\frac{{\bf p}^4}{8M^3_d}\right)
\nonumber\\[2mm]
&&+J^0_{ab}({\bf p},{\bf q})\frac{1}{{\bf k}^2}\,j^0({\bf p},{\bf q})
\nonumber\\[2mm]
&&+J^\alpha_{ab}({\bf p},{\bf q})\frac{1}{{\bf k}^2}\,
\biggl(\delta_{\alpha\beta}
-\frac{k_\alpha k_\beta}{{\bf k}^2}\biggr)j^\beta({\bf p},{\bf q})
\nonumber\\[2mm]
&&+\delta_{ab}V^{\rm vac}({\bf p},{\bf q})\, ,
\en
where ${\bf k}={\bf p}-{\bf q}$ and $a,b=1,2,3$ 
denote the polarization projection of the deuteron. Further, 
$V^{\rm vac}$ stands for the vacuum polarization contribution and
$J^\mu_{ab}$, $j^\mu$ where $\mu=0,\alpha$
denote the electromagnetic formfactor of the deuteron 
and the pion, respectively. At the order of accuracy we are working, one may 
use the following expression for these formfactors

\eq
J^0_{ab}({\bf p},{\bf q})&\!=\!&
e\biggl\{\delta_{ab}\biggl(1-\frac{1}{6}\left<r^ 2_{d}\right>{\bf k}^2\biggr)
+\frac{\mu_Q}{2}\,(k_a k_b - \frac{1}{3}{\bf k}^2\delta_{ab}) 
\nonumber\\[2mm]
&\!+\!& \frac{1}{4M^{2}_{d}}(Q_a k_b - k_a Q_b)(1-\bar\mu_{M})\biggr\} 
+O(\frac{1}{M_d^3})\, , 
\nonumber\\[2mm]
J^\alpha_{ab}({\bf p},{\bf q})&\!=\!&
\frac{e}{2M_d}\left\{Q^\alpha \delta_{ab} +
  \bar{\mu}_{M}(\delta^{\alpha}_{a}k_b - k_a \delta^{\alpha}_{b})\right\}+
O(\frac{1}{M_d^2})\, ,
\nonumber\\[2mm]
j^0({\bf p},{\bf q})&\!=\!&-e\biggl(1
-\frac{1}{6}\left<r^ 2_{\pi}\right>{\bf k}^2\biggr)+O(\frac{1}{M_\pi^3})\, ,
\nonumber\\[2mm]
j^\alpha({\bf p},{\bf q})&\!=\!&-\frac{e}{2M_\pi}\,Q^\alpha+O(\frac{1}{M_\pi^2})\, ,
\en
\begin{sloppypar}
\noindent
where ${\bf Q}={\bf p}+{\bf q}$, and $\mu_Q=0.2859~{\rm fm}^2$,
$\mu_M=\frac{e}{2M_d}\bar\mu_M=0.85741\frac{e}{2m}$ are the deuteron
quadrupole and magnetic moments, respectively. Substituting these relations in
Eq.~(\ref{eq:Vem}), we finally obtain
\end{sloppypar}
\eq\label{eq:Vem_fin}
V^{\rm em}_{ab}({\bf p},{\bf q})&=&-\frac{e^2}{{\bf k}^2}\,\delta_{ab}
+\delta V^{\rm em}_{ab}({\bf p},{\bf q})\, ,
\nonumber\\[2mm]
\delta V^{\rm em}_{ab}({\bf p},{\bf q})
&=&(2\pi)^3\delta^3({\bf q}-{\bf p})\delta_{ab}
\left(\frac{{\bf p}^4}{8M^3_\pi} +\frac{{\bf p}^4}{8M^3_d}\right)
\nonumber\\[2mm]
&-&\frac{e^2}{{\bf k}^2}\biggl\{-\frac{{\bf k}^2}{6}\,\delta_{ab}
\left(\left<r^ 2_{d}\right> +
  \left<r^ 2_{\pi}\right>\right) 
\nonumber\\[2mm]
&+& \frac{1}{4M_d M_\pi}
\left({\bf Q}^2 -\frac{({\bf Q}\cdot{\bf k})^2}{{\bf   k^2}}\right)\delta_{ab} 
\nonumber\\[2mm]
&+&\frac{\mu_Q}{2}\left(k_a k_b -
  \frac{1}{3}{\bf k}^2\delta_{ab}\right) 
\nonumber\\[2mm]
&+& \left(\frac{1-\bar\mu_M}{4M^{2}_{d}}
+ \frac{\bar\mu_M}{2M_dM_\pi}\right)(Q_a k_b - k_a Q_b)\biggr\}
\nonumber\\[2mm]
&+&\delta_{ab}V^{\rm vac}({\bf p},{\bf q})\, .
\en
The electromagnetic energy shift at next-to-leading order is given by
\eq\label{eq:shift}
\Delta E_{nlj}^{\rm em}&=&\sum_{\sigma\rho=\pm,0}\int\frac{d^3{\bf p}}{(2\pi)^3}\,
\frac{d^3{\bf q}}{(2\pi)^3}\,\langle j\nu|l(\nu-\sigma)1\sigma\rangle
\nonumber\\[2mm]
&\times&\langle j\nu|l(\nu-\rho)1\rho\rangle\, 
Y^*_{l(\nu-\sigma)}({\bf p})\chi^*_a(\sigma)
\Psi_{nl}^*(|{\bf p}|)
\nonumber\\[2mm]
&\times&\delta V_{ab}^{\rm em}({\bf p},{\bf q})
\Psi_{nl}(|{\bf q}|)\chi_b(\rho)Y_{l(\nu-\rho)}({\bf q})\, ,
\en
where $Y_{l\nu}$ and $\langle jm|l(m-\sigma)1\sigma\rangle$ are the 
spherical functions and the Clebsch-Gordan coefficients, respectively and
\eq\label{eq:chi}
\chi(+)=\frac{1}{\sqrt{2}}\begin{pmatrix}{-1\cr -i\cr 0}\end{pmatrix} ,\,
\chi(-)=\frac{1}{\sqrt{2}}\begin{pmatrix}{1\cr -i\cr 0}\end{pmatrix},\,
\chi(0)=\begin{pmatrix}{0\cr 0\cr 1}\end{pmatrix} .
\nonumber\\
\en
Note that the energy shift (\ref{eq:shift}) does not depend on the magnetic 
quantum number $\nu$.
Substituting here the expression (\ref{eq:Vem_fin}), and calculating the 
integral, we finally obtain
\eq
\Delta E_{nlj}^{\rm em}&=&-\frac{M_d^3+M_\pi^3}{8M_d^3M_\pi^3}\,
\biggl(\frac{\alpha\mu_d}{n}\biggr)^4\biggl\{\frac{4n}{l+1/2}-3\biggr\}
\nonumber\\[2mm]
&+&\delta_{l0}\,
\frac{2}{3n^3}\,\alpha^4\mu_d^3(\langle r_d^2\rangle+\langle r_\pi^2\rangle)
\nonumber\\[2mm]
&-&\frac{\alpha^4\mu_d^3}{4M_dM_\pi n^4}\biggl\{-4n\delta_{l0}-4
+\frac{6n}{l+1/2}\biggr\}
\nonumber\\[2mm]
&+&\Delta E_{nlj}^{\rm Q}+\Delta E_{nlj}^{\rm M}+\Delta E_{nlj}^{\rm vac}\, ,
\en
where for $l\neq 0$
\eq
\Delta E^{Q}_{nlj} &=& (-1)^{(l+j+1)}\,\frac{2\alpha^4
  \mu^{3}_{d}}{n^3}\,\mu_Q\,\sqrt{\frac{30(2l-2)!}{(2l+3)!}}
\nonumber\\
&\times&\left\{
  \begin{array}{ccc}
l & 1 & j \\
1 & l & 2 \\
\end{array}\right\} \,,\\
\Delta E^{M}_{nlj}&=&(-1)^{(l+j)}\,\frac{\alpha^4
  \mu^{3}_{d}}{n^3}\,2\sqrt{6}\left(\frac{1-\bar\mu_M}{2M^2_d} 
+ \frac{\bar\mu_M}{M_d M_\pi}\right)
\nonumber\\
&\times&\frac{\left\{
  \begin{array}{ccc}
l & 1 & l \\
1 & j & 1 \\
\end{array}\right\}}{\sqrt{l(l+1)(2l+1)}}\, ,
\en
and for $l=0$, $\Delta E^{Q}_{nlj} =\Delta E^{M}_{nlj}= 0$. The quantities
in the braces denote the Wigner $6j$ symbols. Finally, the vacuum 
polarization contribution is given in Eq.~(3) of Ref.~\cite{Eiras}.
We do not display it here.

Taking $l=1$, we have
\eq
\Delta E^{Q}_{n1j}&=&\frac{\alpha^4 \mu^3_{d}\mu_Q}{3n^3}\left(
\begin{array}{r}
1\,\,\,\,;\,\,\,\,j=0 \\
 \\
-\frac{1}{2}\,\,\,\,;\,\,\,\,j=1  \\
\\
\frac{1}{10}\,\,\,\,;\,\,\,\,j=2\\
\end{array}\right)
\nonumber\\[2mm]
&=&\frac{0.015~{\rm eV}}{n^3}\left(
\begin{array}{r}
1\,\,\,\,;\,\,\,\,j=0 \\
 \\
-\frac{1}{2}\,\,\,\,;\,\,\,\,j=1  \\
\\
\frac{1}{10}\,\,\,\,;\,\,\,\,j=2\\
\end{array}\right)
\en
\eq
\Delta E^{M}_{n1j} &=& \frac{2\alpha^4 \mu^3_{d}}{3n^3}\left(
\frac{1-\bar\mu_M}{2M^2_d} + \frac{\bar\mu_M}{M_d M_\pi}\right)
\left(
\begin{array}{r}
1\,\,\,\,;\,\,\,\,j=0 \\
 \\
\frac{1}{2}\,\,\,\,;\,\,\,\,j=1  \\
\\
-\frac{1}{2}\,\,\,\,;\,\,\,\,j=2\\
\end{array}\right)
\nonumber\\[2mm]
&=&\frac{0.027~{\rm eV}}{n^3}\left(
\begin{array}{r}
1\,\,\,\,;\,\,\,\,j=0 \\
 \\
\frac{1}{2}\,\,\,\,;\,\,\,\,j=1  \\
\\
-\frac{1}{2}\,\,\,\,;\,\,\,\,j=2\\
\end{array}\right)
\en
As we see, the splitting of the energy levels is tiny. Note that 
in the averaged level energy, defined by Eq.~(\ref{eq:E-average}), 
these contributions vanish:
$\overline{\Delta E_{nlj}^{Q}}=\overline{\Delta E_{nlj}^{M}}=0$.
The remaining contributions can be rearranged in order to allow the comparison
with the results existing in the literature.
In this manner, we finally arrive at the result given in Eqs. 
(\ref{eq:separation}) and (\ref{eq:em}).

\setcounter{equation}{0}
\section{Table of integrals}
\label{sec:int}

After integrating over the 0th components $l^0$, $k^0$, the quantity $R_a$ 
in Eq.~(\ref{eq:NLO}) can be rewritten as
\eq\label{eq:b1}
R_a&=&\int \frac{d^d{\bf l}}{(2\pi)^d}\,
\frac{d^d{\bf k}}{(2\pi)^d}\,
\frac{2m^3(d_+^2-2d_-^2)}{(\gamma^2+{\bf l}^2)(\gamma^2+{\bf k}^2)}
\nonumber\\[2mm]
&\times&\frac{1}{
\gamma^2+\frac{1}{2}\,[{\bf l}^2+{\bf k}^2+\frac{m}{M_\pi}({\bf l}-{\bf k})^2]}
\, .
\en
Introducing Feynman parameters and carrying out the integration over the 
momenta in the standard manner, one gets
 \eq\label{eq:b2}
R_a&=&2m^3(d_+^2-2d_-^2)\,\frac{\Gamma(3-d)\mu^{2(d-3)}}{(4\pi)^d}\,
\biggl(\frac{\gamma^2}{\mu^2}\biggr)^{d-3}
\nonumber\\[2mm]
&\times&\int dx_1dx_2dx_3\delta(1-x_1-x_2-x_3)[G_a]^{-\frac{d}{2}}\, ,
\nonumber\\[2mm]
G_a&=&x_1x_2+\frac{x_3}{2}\,\biggl(1-\frac{x_3}{2}+\frac{m}{M_\pi}\biggr)\, ,
\en
where $\mu$ denotes the scale of  dimensional regularization.
Renormalizing $R_a$ according to the $\overline{\rm MS}$ prescription, 
for the finite part we get
\eq\label{eq:b3}
R_a^{\rm\, fin}&\!=\!&\frac{m^3(d_+^2-2d_-^2)}{32\pi^3}\,\biggl(
x_a\ln\frac{\gamma^2}{\mu^2}+J_a\biggr)+O(d-3)\, ,
\nonumber\\[2mm]
x_a&\!=\!&-\int dx_1dx_2dx_3\delta(1-x_1-x_2-x_3)[G_a]^{-\frac{3}{2}}
\nonumber\\[2mm]
&\!=\!&-\frac{8M_\pi}{m}\,\arcsin\frac{m}{m+M_\pi}=-1.2569\cdots\, ,
\nonumber\\[2mm]
J_a&\!=\!&\frac{1}{2}\,
\int dx_1dx_2dx_3\delta(1-x_1-x_2-x_3)[G_a]^{-\frac{3}{2}}\ln[G_a]
\nonumber\\[2mm]
&\!=\!&-1.1334\cdots\, .
\en
In the same manner, one obtains
\eq\label{eq:b4}
R_b&=&2m^3(d_+^2+2d_-^2)\,\frac{\Gamma(3-d)\mu^{2(d-3)}}{(4\pi)^d}\,
\nonumber\\[2mm]
&\times&\int_0^1 dx(1-x)\biggl(\frac{xm}{2\mu_N}\biggr)^{-\frac{d}{2}}
\biggl(1-\frac{\mu_N x}{2m}\biggr)^{-\frac{d}{2}}\, ,
\en
where $\mu_N=mM_\pi(m+M_\pi)^{-1}$ is the reduced mass in the $\pi N$
system. After the renormalization one may evaluate the remaining integrals 
analytically. As the result, one obtains
\eq\label{eq:b5}
R_b^{\rm\, fin}&\!=\!&\frac{m^3(d_+^2+2d_-^2)}{32\pi^3}\,\biggl(
x_b\ln\frac{\gamma^2}{\mu^2}+J_b\biggr)+O(d-3)\, ,
\nonumber\\[2mm]
x_b&\!=\!&32\kappa^{\frac{3}{2}}\sqrt{1-\kappa}=0.5098+\cdots
\nonumber\\[2mm]
J_b&\!=\!&16\kappa^{\frac{3}{2}}\biggl\{
\frac{2}{\sqrt{\kappa}}(1-2\kappa)\arcsin\sqrt{\kappa}
\nonumber\\[2mm]
&\!-\!&\sqrt{1-\kappa}\,
(2+\ln(1-\kappa)-\ln4\kappa)\biggr\}
\nonumber\\[2mm]
&\!=\!&-0.3731+\cdots\, ,\quad\quad
\kappa=\frac{\mu_N}{2m}\, .
\en
The contribution $R_c$ can be evaluated in a similar manner. We do not 
display this calculation here, since this contribution is proportional to
$a_+^2$ and drops anyway. The same method can be applied to the calculation of the diagrams contributing to the threshold scattering amplitude 
${\cal A}(\mu)$.

\setcounter{equation}{0}
\section{Resonance saturation for the LEC $f_0$}
\label{sec:resonance}

\begin{figure}[t]
\begin{center}
\includegraphics[width=6.cm]{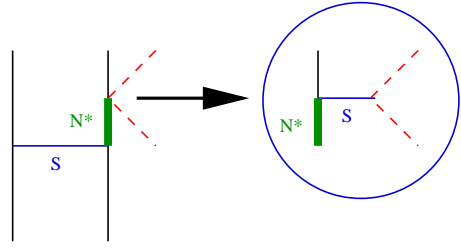}
\end{center}
\caption{Tree-level contribution to the scattering amplitude $\pi NN\to \pi NN$
with the exchange of the $N^*(1440)$ and the scalar meson.}
\label{fig:Nstar}
\end{figure}

In order to get an estimate for the LEC $f_0'$, we have calculated the tree
amplitude of the process $\pi NN\to \pi NN$ in the theory with the explicit
$\Delta$, $N^*(1440)$, scalar and vector mesons in the limit when masses
of all above resonances become very large. At threshold, the $\Delta$ and the 
vector meson do not contribute\footnote{Vector mesons are described by 
antisymmetric tensor fields.}. In Fig.~\ref{fig:Nstar} we display an example
of a diagram that does not vanish at threshold. One has 4 such diagrams
(insertion of $N^*(1440)$ 
in each nucleon leg), as well as the diagrams which are 
obtained by a permutation of the outgoing nucleon legs.

The Lagrangian describing the interactions between nucleons, pions, scalar 
mesons and the $N^*(1440)$ is given by (see also \cite{BKMlec})
\eq\label{eq:lagr_C}
{\cal L}^*&=&g_{sNN}S\bar NN+[g_{sNN^*}S\bar N^*N+{\rm h.c.}]
\nonumber\\[2mm]
&+&[c_1^*\bar N^*\chi_+N-\frac{c_2^*}{m^{*\,2}}\,(\partial_\mu\partial_\nu
\bar N^*)u^\mu u^\nu N+{\rm h.c.}]\, ,\hspace*{1.cm}
\en
where $\chi_+=M_\pi^2(2-{\mathbold{\pi}^2}/F^2+\cdots)$ and
$u_\mu=-\mathbold{\tau}\partial_\mu\mathbold{\pi}/F+\cdots$, and $m^*$ denotes
the mass of the $N^*(1440)$. The contribution to the 
scattering amplitude at threshold in the pertinent spin-isospin channel 
from the diagram in Fig.~\ref{fig:Nstar}
is given by
\eq\label{eq:res}
T^*&=&-(2m)^2\frac{2M_\pi^2(c_1^*-c_2^*)g_{sNN}g_{sNN^*}}{F^2m^*M_S^2}\,
\nonumber\\[2mm]
&\times&(\chi_1^\dagger\chi_1\chi_2^\dagger\chi_2+
\chi_1^\dagger\chi_2\chi_2^\dagger\chi_1)\, ,
\en
where $M_S$ stands for the mass of the scalar meson. From the above 
expression, one may read off the contribution to the LEC $f_0'$
\eq\label{eq:nS}
&&f_0'=\frac{\xi}{2F^4M_\pi} \biggl(\frac{M_\pi}{4\pi F}\biggr)^2\, ,
\nonumber\\[2mm]
&&\xi=-\frac{64\pi^2 F^4(c_1^*-c_2^*)g_{sNN}g_{sNN^*}}{m^*M_S^2}\, ,
\en
\begin{sloppypar}
\noindent
where, according to the natural size arguments (see Eq.~(\ref{eq:Friar})),
the quantity $\xi$ must be of order 1. 
\end{sloppypar}

In the numerical estimates, we use the scalar meson mass $M_S=550~{\rm MeV}$
(as it was also done in Ref.~\cite{BKMlec}).
Further, $g_{sNN}$ and $g_{sNN^*}$ denote the values of the pertinent 
formfactors at zero momentum transfer. With the choice of monopole formfactor
$F_S(q^2)=(\Lambda_S^2-M_S^2)/(\Lambda_S^2-q^2)$ with 
$\Lambda_S=1700~{\rm MeV}$ for both $SNN$ and $SNN^*$ vertices, we
deduce $g_{sNN}=7.57$ and $g_{sNN^*}=3.66$ from the mass-shell values given
e.g. in Ref.~\cite{Ruso}. Further, in the same paper, one finds
two different values for the constants $c_1^*,c_2^*$
\eq\label{eq:c1c2}
\begin{array}{l l}
c_1^*=-7.27~{\rm GeV}^{-1}\, ,\quad c_2^*=0 &\,\,\mbox{[Set 1]}\, ,
\\[2mm]
c_1^*=-12.7~{\rm GeV}^{-1}\, ,\quad
c_1^*=1.98~{\rm GeV}^{-1}&\,\,\mbox{[Set 2]}\, .
\end{array}
\en

Substituting these values, we get $\xi=21$ [Set 1] and $\xi=43$ [Set 2]
that obviously contradicts to our natural-size estimate. Note that the values
for $c_1^*,c_2^*$ given in Ref.~\cite{Ruso} also do not agree with
the result of the fit $c_1^*+c_2^*=(-1.56\pm 3.35)~{\rm GeV}^{-1}$ 
from Ref.~\cite{BKM}. However, in order to use the latter fit in our estimate,
needs one more relation between $c_1^*$ and $c_2^*$. We get this relation,
assuming that these constants are saturated by the scalar meson exchange.
The Lagrangian describing the interaction of pions with the scalar field
with total isospin $I=0$ can be taken from Ref.~\cite{Pich}
\eq\label{eq:pich}
{\cal L}_S=\tilde c_d S\,{\rm Tr}(u_\mu u^\mu)
+\tilde c_m S{\rm Tr}(\chi_+)\, .
\en
from the saturation hypothesis one gets $c_1^*/c_2^*=\tilde c_m/\tilde c_d
=4.2/3.2$ (in the last equation, we have used numerical values from 
Ref.~\cite{Pich}). Together with the value of $c_1^*+c_2^*$ from 
Ref.~\cite{BKM} this yields the estimate $\xi=0.6\pm 1.3$, which is in a 
reasonable agreement with the natural-size conjecture. Note that,
albeit the resonance saturation method at present does not lead to an improved
accuracy as compared to the dimensional analysis, the above discussion
demonstrates the feasibility of estimating $f_0'$ and eventually $f_0$
through  resonance
saturation. What is needed to this end is to determine the LECs
$c_1^*,c_2^*$ more accurately from the fit to $N^*(1440)\to N\pi\pi$ and,
in particular, to resolve the large discrepancies reported in 
Refs.~\cite{BKM,Ruso}.

\newpage

\section*{Erratum}

In the above paper we have studied the pion-deuteron
scattering length within low-energy effective theories of QCD.
In particular, in the section \ref{sec:IF} of this article we give an
estimate of the uncertainty in this scattering length, coming from
the presence of the 6-particle low-energy constant (LEC) -- referred
to as $f_0$. Unfortunately, in this estimate, which has yielded
$\Delta a_{\pi d}/a_{\pi d}\simeq 12\%$, an incorrect value for the
deuteron wave function at the origin in the Weinberg approach
$\tilde \Phi_0(\Lambda)= [0.487;0.434]~{\rm fm}^{-3/2}$
with the cutoff mass
from the interval $\Lambda=[450;650]$ has been used (normalization error).
Using the corrected value 
$\tilde \Phi_0(\Lambda)=[0.137;0.122]~{\rm fm}^{-3/2}$
downsizes the error
 (only due to the above-mentioned source) to 
a much more comfortable $\Delta a_{\pi d}/a_{\pi d}\simeq 1\%$,
which, in addition, fits better to the {\it a priori} estimates, based on the
Weinberg power counting. Thus, the results of the high-precision calculations of
Ref.~\cite{Bernard} have been independently confirmed.

This result has far-reaching consequences. In particular, we have to 
re-think the equivalence between the Weinberg \cite{Bernard} and
HPEFT \cite{Borasoy,Beane} 
approaches to the pion-deuteron scattering at low energy.
The arguments remain the same, but the big change in the estimated
uncertainty leads to the conclusion that now
this equivalence is realized in a slightly different manner.
As before, the success of the modified power counting \cite{Bernard} unambiguously
indicates that HPEFT is a physically equivalent tool for describing 
pion-deuteron interactions near threshold. Furthermore, 
the sole input which HPEFT imports from the 
Weinberg approach are the values of LECs, determined through the matching
procedure. As it is discussed in the above paper, these LECs in 
general acquire contributions from two different momentum regions:
$M_\pi< p<\Lambda$ and $p>\Lambda$. Whereas the former can in principle be
expressed through the wave functions, etc -- and is thus calculable
in terms of the known parameters of the theory, the latter represents a 
genuinely high-energy contribution, which at the present stage can only
be included in the error estimate. What follows from our corrected 
calculations
is that this high-energy contribution is small and does not preclude a
high-precision determination of the pion-nucleon scattering lengths from the
combined analysis of the pionic hydrogen and pionic deuterium data by
Pionic Hydrogen collaboration at PSI \cite{PSI1,PSI2}. We wish to emphasize once
more that the strong predictive power is a result of a subtle balance 
between HPEFT, which at the end allows for constructing multiple
scattering series in terms of observables (scattering lengths, etc)
and the Weinberg approach, which enables one to evaluate the LECs of 
HPEFT with a high precision.

Several remarks are in order. First, we wish to mention that our error 
estimate, which is carried out by using dimensional arguments or resonance 
saturation hypothesis, is complementary to the study of the scale dependence
of the calculated pion-deuteron scattering length. 
Obviously, the investigations should proceed from both ends.
As mentioned in the above paper, a weak scale
dependence would indicate that the uncertainty might be small and not 
that it must be 
small.  Finally, the fact that the short-range interactions do not introduce
large uncertainty in the calculated value of the pion-deuteron scattering 
length, might stimulate further activity in calculating various small
contributions, with an aim to determine this scattering length as precise
as possible. 
Using the above-mentioned equivalence
between the Weinberg and HPEFT app\-ro\-a\-ches could enable one to carry out
these calculations in a simpler setting.

\end{document}